\def\3cm{\rm {cm^{-3}}}
\def\2cm{\rm {cm^{-2}}}
\def\s-1{\rm {s^{-1}}}
\def\etal {et al.}
\def\kms {\hbox{${\rm km\,s}^{-1}$}}
\def\Kkms {\hbox{${\rm K}\,{\rm km}^{-1}\,{\rm s}$ }}
\def\ndv{\hbox{${\rm cm}^{-2}\,{\rm km}^{-1}\,{\rm s} $}}

\def\twco{$^{12}$CO~}
\def\thco{$^{13}$CO~}
\def\oi{[O I]~}
\def\ci{[C I]~}
\def\cii{[C II]~}
\def\c18o{C$^{18}$O~}

\documentclass{aa}  
\usepackage{multirow}
\usepackage{graphicx}
\usepackage{epsfig}
\usepackage{txfonts}
\usepackage{psfrag}
\usepackage{units}
\begin{document}
\title{CHAMP$^+$ observations of warm gas in M17 SW
}
\subtitle{}
\author{J.P. P\'{e}rez-Beaupuits\inst{1} \and 
        M.~Spaans\inst{1} \and	
	M.R.~Hogerheijde\inst{2} \and	
        R.~G\"usten\inst{3} \and
        A.~Baryshev\inst{4} \and 
        W.~Boland\inst{2,5}
}
\offprints{J.P. P\'erez-Beaupuits}
\institute{
 Kapteyn Astronomical Institute, Rijksuniversiteit Groningen, 9747 AV Groningen, The Netherlands - 
 \email{jp@astro.rug.nl}
 \and
 Leiden Observatory, Leiden University, PO Box 9513, 2300 RA, Leiden, The Netherlands 
 \and
 Max-Planck-Institut f\"ur Radioastronomie, Auf dem H\"ugel 69, 53121 Bonn, Germany 
 \and
 SRON Netherlands Institute for Space Research , PO Box 800, 9700 AV Groningen, The Netherlands
 \and
 Nederlandse Onderzoeksschool Voor Astronomie (NOVA), PO Box 9513, 2300 RA Leiden, The Netherlands 
}
\date{Received  / Accepted  }
\titlerunning{Warm gas in M17 SW}

%
\abstract
  {Sub-millimeter and Far-IR observations have shown the presence of a significant amount of warm (few hundred K) and dense ($n(\rm H_2)\ge10^4~\3cm$) gas in sources ranging from active star forming regions to the vicinity of the Galactic center. Since the main cooling lines of the gas phase are important tracers of the interstellar medium in Galactic and extragalactic sources, proper and detailed understanding of their emission, and the ambient conditions of the emitting gas, is necessary for a robust interpretation of the observations.
  }
  {With high resolution ($7''-9''$) maps ($\sim3\times3$ pc$^2$) of mid-$J$ molecular lines we aim to probe the physical conditions and spatial distribution of the warm (50 to few hundred K) and dense gas ($n(\rm H_2)>10^5~\3cm$) across the interface region of the nearly edge-on M17 SW nebula. 
  }
  {We have used the dual color multiple pixel receiver CHAMP$^+$ on APEX telescope to obtain a $5'.3\times4'.7$
  map of the $J=6\rightarrow5$ and $J=7\rightarrow6$ transitions of \twco, the \thco $J=6\rightarrow5$ line,
  and the $^3P_2\rightarrow{^3P_1}$ 370 $\mu$m fine-structure transition of \ci in M17 SW. LTE and non-LTE radiative transfer models are used to constrain the ambient conditions.
  }
  {The warm gas extends up to a distance of $\sim2.2$ pc from the M17 SW ridge. The \thco\ $J=6\rightarrow5$ and \ci\ 370 $\mu$m lines have a narrower spatial extent of about 1.3 pc along a strip line at P.A=$63^{\circ}$. The structure and distribution of the \ci $^3P_2\rightarrow{^3P_1}$ 370 $\mu$m map indicate that its emission arises from the interclump medium with densities of the order of $10^3~\3cm$.
  }
  {The warmest gas is located along the ridge of the cloud, close to the ionization front. An LTE approximation indicates that the excitation temperature of the embedded clumps goes up to $\sim120$ K. The non-LTE model suggests that the kinetic temperature at four selected positions cannot exceed 230 K in clumps of density $n(\rm H_2)\sim5\times10^5~\3cm$, and that the warm ($T_k>100$ K) and dense ($n(\rm H_2)\ge10^4~\3cm$) gas traced by the mid-$J$ \twco lines represent just about 2\% of the bulk of the molecular gas. The clump volume filling factor ranges between 0.04 and 0.11 at these positions.
  }
  

\maketitle

\section{Introduction}

The heating and cooling balance in photon-dominated regions (PDRs) remains an active study of research.
The comprehensive understanding of PDRs requires observations of large areas close to radiation sources, 
and of a wide wavelength range covering various emissions of atoms, molecules, and grains.
In particular, mid-$J$ CO lines have been detected in almost all known massive Galactic star forming regions (e.g. Orion Nebula, W51, Cepheus A, NGC 2024). This indicates that warm ($T_K\ge50$ K) and dense ($n(H_2)\ge10^4~\3cm$) gas is common, and probably of importance in most OB star forming regions. The mid-$J$ CO lines detected in regions like, e.g. M17, Cepheus A and W51, have relatively narrow line widths of 5--10 \kms, although not as narrow as the line widths observed in cold quiescent cloud cores. 

Observations of the $J=6\rightarrow5$ and $J=7\rightarrow6$ transitions of \twco in several massive star forming regions indicate that the warm emitting gas is confined to narrow ($<1$ pc) zones close to the ionization front. These observations favor photoelectric heating of the warm gas by UV radiation fields outside the HII regions (e.g. Harris \etal\ 1987; Graf \etal\ 1993; Yamamoto \etal\ 2001; Kramer \etal\ 2004 and 2008). Nevertheless, shocks may also be an important source of heating in high velocity wing sources like, Orion, W51 and W49 (Jaffe \etal\ 1987).
 
Because of its nearly edge-on geometry, and the large amount of observational 
data available in the literature, M17 SW is one of the best 
Galactic regions to study the entire structure of PDRs from the 
exciting sources to the ionization front, and the succession (or not) of 
H$_2$, \ci and CO emissions, as predicted by PDR models (Icke, Gatley, \& Israel 1980; Felli, Churchwell, \& Massi 1984; Meixner et al. 1992; Meijerink \& Spaans 2005). M17 SW is also one of the few star forming regions for which the magnetic field strength can be measured in the PDR interface, and where the structure of the neutral and molecular gas seems to be dominated by magnetic pressure rather than by gas pressure (Pellegrini \etal\ 2007).

M17 SW is a giant molecular cloud at a distance of 2.2 kpc,
illuminated by a highly obscured ($A_v>10$ mag) cluster of several OB stars (among $\ga100$ stars) 
at about 1 pc to the East (Beetz \etal\ 1976; Hanson \etal\ 1997). It also harbours a number of candidate young stellar objects that have recently been found (Povich \etal\ 2009).
Several studies of molecular emission, excitation and line 
profiles (e.g. Snell \etal~1984; Martin, Sanders \& Hills 1984; 
Stutzki \& G\"usten 1990) from the M17 SW core indicate that 
the structure of the gas is highly clumped rather than homogeneous.
Emission of \ci and \cii was detected more than a parsec into
the molecular cloud along cuts through the interface region (Keene \etal\ 1985; Genzel \etal\ 1988; Stutzki \etal\ 1988). 
These results, as well as those found in other star forming regions like S106, the Orion Molecular Cloud, and the NGC~7023 Nebula (e.g. Gerin \& Phillips 1998; Yamamoto \etal\ 2001; Schneider \etal\ 2002, 2003; Mookerjea \etal\ 2003) do not agree with the atomic and molecular stratification predicted by standard steady-state PDR models.
However, the extended \ci $^3P_1\rightarrow{^3P_0}$ and \thco $J=2\rightarrow1$ emission in S140 have been successfully explained by a stationary, but clumpy, PDR model (Spaans 1996; Spaans \& van Dishoeck 1997).
Hence, the lack of stratification in \ci, \cii and CO is a result
that can be expected for inhomogeneous clouds, where each clump acts
as an individual PDR. On the other hand, a partial face-on illumination of the molecular clouds would also suppress stratification.

Based on analysis of low-$J$ lines of \twco, \thco and CH$_3$CCH data, 
the temperature towards the M17 SW cloud core has been estimated as 50--60 K, 
whereas the mean cloud temperature has been found to be about 30--35 K
(e.g. G\"usten \& Fiebig 1988; Bergin \etal\ 1994; Wilson \etal\ 1999; Howe \etal\ 2000; Snell \etal\ 2000).
Temperatures of $\sim275$ K has been estimated from NH$_3$ observations (G\"usten \& Fiebig 1988) towards the VLA continuum arc, which agree with estimates from highly excited \twco transitions (Harris \etal\ 1987).
Multitransition CS and HC$_3$N observations indicates that the density at the core region
of M17 SW is about $6\times10^5~\3cm$ 
(e.g., Snell \etal\ 1984; Wang \etal\ 1993; Bergin, Snell \& Goldsmith 1996). 
While densities up to $3\times10^6~\3cm$ have been estimated towards the north rim with multitransition observations of NH$_3$, which indicates that ammonia is coexistent with high density material traced in CS and HCN (G\"usten \& Fiebig 1988).
The UV radiation field $G_0$ has been estimated to be of the order of $10^4$ in units of the ambient interstellar radiation field ($1.2\times10^{-4}~\rm ergs~s^{-1}~cm^{-1}~sr^{-1}$, Habing 1968; Meixner \etal\ 1992).


However, most of the millimeter-wave molecular observations in M17 SW are sensitive only to low temperatures ($<100$ K),
and the few available data of mid-$J$ CO and \ci lines (consisting mostly of cuts across the ionization front and observations at few selected positions) are limited in spatial resolution and extent (e.g. Harris \etal\ 1987; Stutzki \etal\ 1988; Genzel \etal\ 1988; Stutzki \& G\"usten 1990; Meixner \etal\ 1992; Graf \etal\ 1993; Howe \etal\ 2000). Therefore, in this work we present maps ($\sim3\times3$ pc$^2$) of mid-$J$ molecular (\twco and \thco) and atomic (\ci) gas, with excellent
high resolution ($9.4''-7.7''$), which advances existing work in M17 SW.

The observations were done with CHAMP$^+$ (Carbon Heterodyne Array of the MPIfR) on the Atacama Pathfinder EXperiment (APEX\footnote{This publication is based on data acquired with the Atacama Pathfinder Experiment (APEX). APEX is a collaboration between the Max-Planck-Institut fur Radioastronomie, the European Southern Observatory, and the Onsala Space Observatory}) (G\"usten \etal\ 2006). The multiple pixels at two submm frequencies of CHAMP+, allow for efficient mapping of $\sim$arcmin regions, and provide the ability to observe simultaneously the emission from the $J=6\rightarrow5$ and $J=7\rightarrow6$ rotational transitions of \twco at 691.473 GHz and 806.652 GHz, respectively.
We also observed the $J=6\rightarrow5$ transition of \thco at 661.067 GHz and the $^3P_2\rightarrow~^3P_1$ 370 $\mu$m (hereafter: $2\rightarrow1$) fine-structure transition of \ci at 809.342 GHz.

Since the gas phase cools mainly via the atomic fine structure lines of \oi, \cii, \ci, and the rotational CO lines (e.g. Kaufman \etal\ 1999, Meijerink \& Spaans 2005), these carbon bearing species presented here are very important coolants in the interstellar medium (ISM) of a variety of sources in the Universe, from Galactic star forming regions, the Milky Way as a galaxy, and external galaxies up to high redshifts (e.g. Fixsen \etal\ 1999; Weiss \etal\ 2003; Kramer \etal\ 2005; Bayet \etal\ 2006; Jakob \etal\ 2007).

The case of M17 SW can be considered as a proxy for extra galactic star forming regions. 
M17 SW is not special, nor does it need to be, compared to other
massive star-forming regions like Orion, W49, Cepheus A, or W51. Still, it does allow
feedback effects, expected to be important for starburst and active galaxies, to be
studied in great spatial detail. Comparison of the local line ratios to
extra-galactic regions can then shed light on the properties of massive
star forming regions that drive the energetics of active galaxies.
Our results will be of great use for future high resolution observations, since molecular clouds of the size of the maps we present will be resolved by ALMA at the distance ($\sim14$ Mpc) of galaxies like NGC~1068.

The main purpose of this work is to explore the actual spatial distribution of the mid-$J$ \twco\ and \ci lines in M17 SW, and to test the ambient conditions of the warm gas.
A simple LTE model based on the ratio between the \twco and \thco $J=6\rightarrow5$ lines is 
used to probe the temperature of the warm ($T_K\sim100$ K) and dense ($n_{\rm H}>10^5~\3cm$) molecular gas.
Then a non-LTE model is used to test the ambient conditions at four selected positions.
In a follow up work we will present an elaborate model of these high resolution data.

The most frequent references to Stutzki \etal\ (1988), Stutzki \& G\"usten (1990) and Meixner \etal\ (1992) will be referred to as S88, SG90 and M92, respectively. The organization of this article is as follows. In \S2 we describe the observations. The maps of the four lines observed are presented in \S3. The modelling and analysis of the ambient conditions are presented in \S4. And the conclusions and final remarks are presented in \S5.


\section{Observations}

We have used the dual color heterodyne array receiver CHAMP$^+$ (Kasemann \etal\ 2006; G\"usten \etal\ 2008), providing $2\times7$ pixels, 
on the APEX telescope during July 2008, to map simultaneously the $J = 6\rightarrow5$ and $J = 7\rightarrow6$ 
lines of \twco, and - in a second coverage - the \thco $J = 6\rightarrow5$ and \ci $J = 2\rightarrow1$.
We observed a region of about $5'.3\times4'.7$ (3.4 pc $\times$ 3.0 pc) in on-the-fly (OTF) 
slews in R.A. ($\sim320$ arcsec long), subsequent scans spaced by $4''$ in Declination.
The observations were done in total power mode (nodding the antenna prior to each OTF 
slew to a reference position $180''$ east of the SAO star 161357. The latter is used as 
reference throughout the paper, with R.A(J2000)=18:20:27.64 and Dec(J2000)=-16:12:00.90. 
We used Sgr B2(N) as reference for continuum poiting.
Calibration measurements were performed regularly, every $\sim$10 min, with a cold liquid nitrogen (LN2) load and an ambient temperature load. The data were processed with the APEX real-time calibration software (Muders \etal\ 2006), assuming an image sideband suppression of 10 dB.

\begin{figure}[!ht]
  
  \hfill\includegraphics[width=7cm,angle=-90]{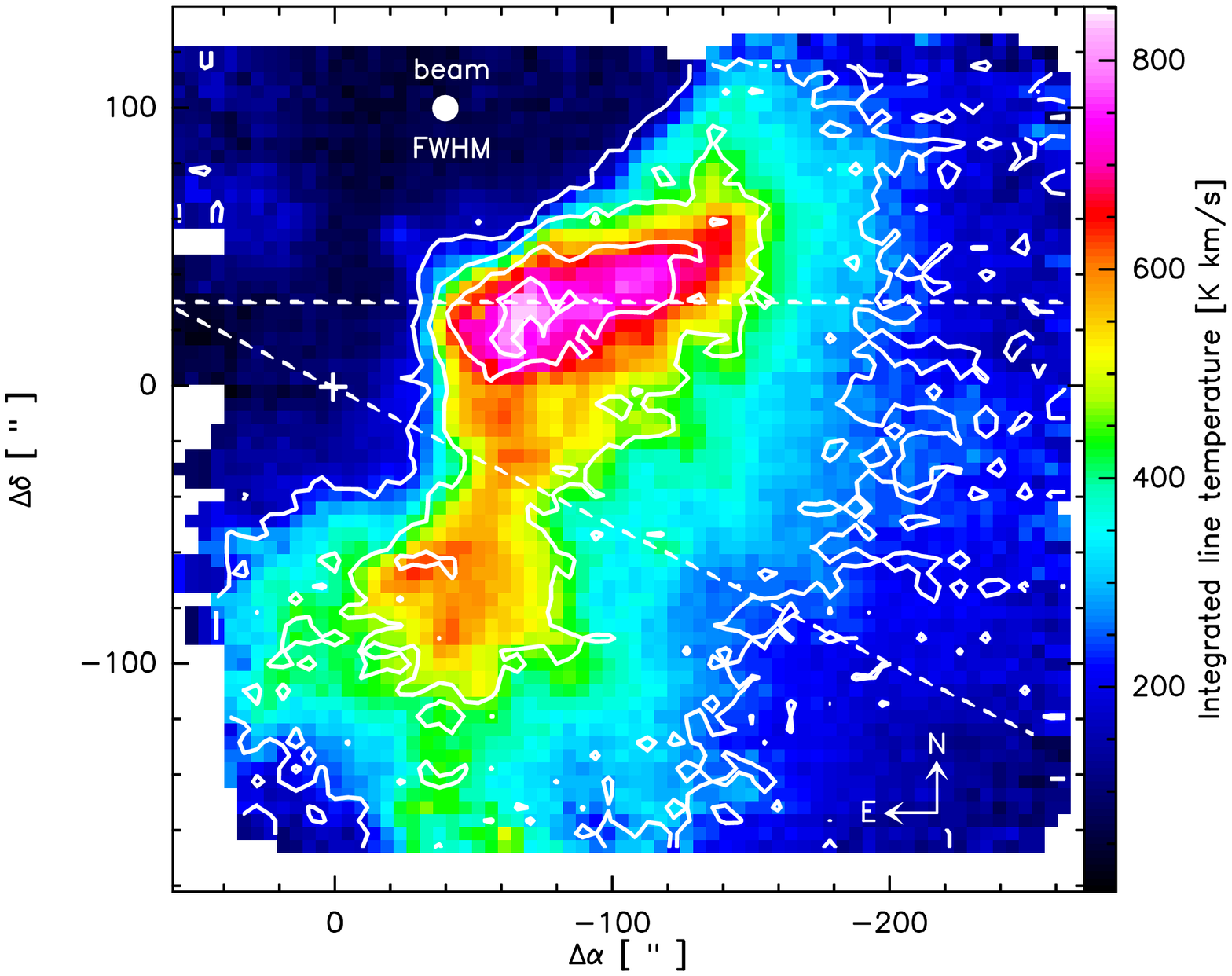}\hspace*{\fill}\\
  \hfill\includegraphics[width=7cm,angle=-90]{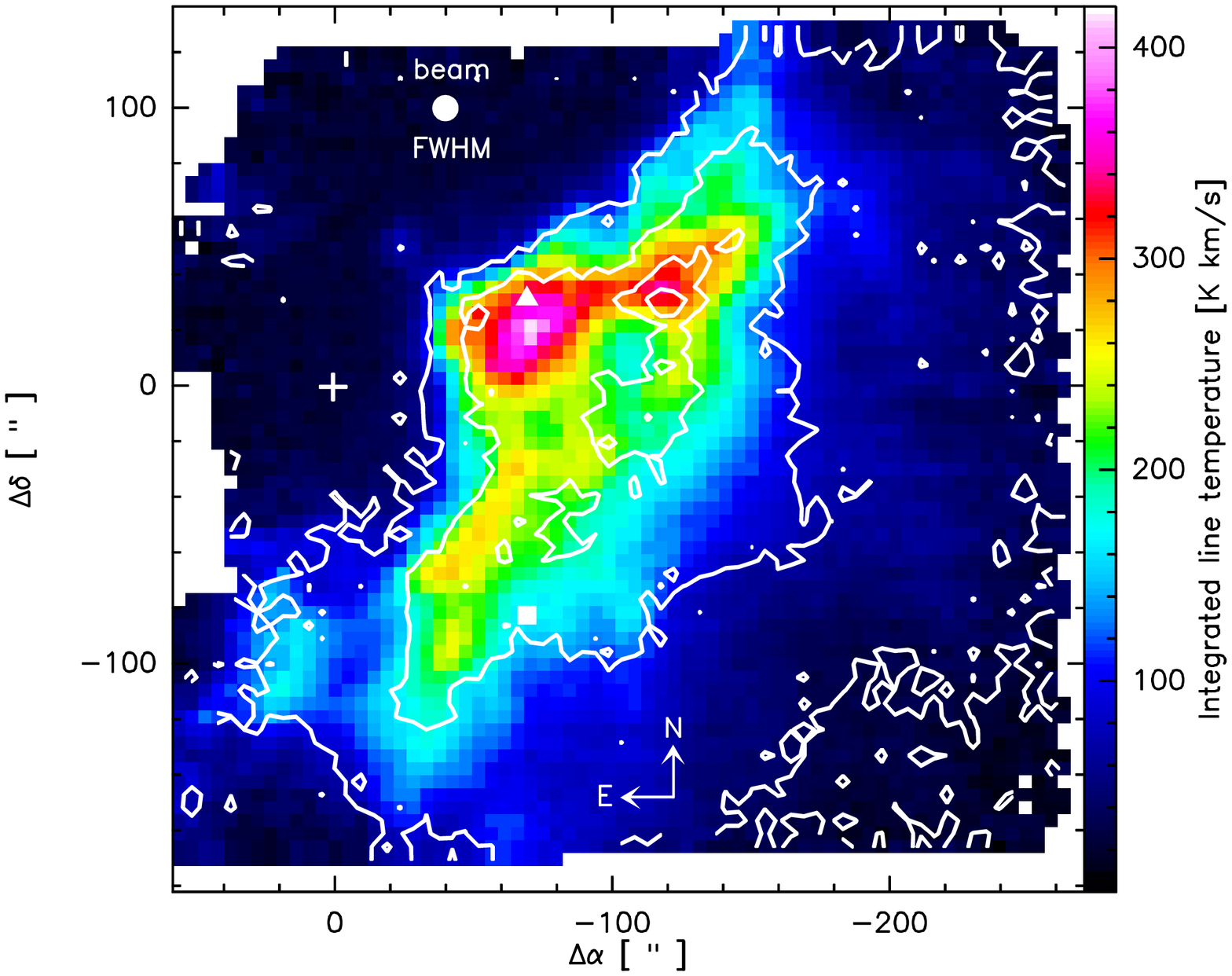}\hspace*{\fill}

  \caption{\footnotesize{\textit{Top} - Color map of the integrated temperature of \twco $J = 6\rightarrow5$ in M17 SW. The contour lines correspond to the \twco $J = 7\rightarrow6$, which has a peak emission of 925 K \kms. The countour levels are the 25\%, 50\%, 75\% and 90\% of the peak emission. Dashed lines correspond to the E-W and NE-SW strip lines at P.A=$90^{\circ}$ and P.A=$63^{\circ}$, respectively. \textit{Bottom} - Color map of the integrated temperature of \thco $J = 6\rightarrow5$ and the contour levels (as described before) of \ci\ $J = 2\rightarrow1$ with a peak emission of 282 K \kms. The filled \textit{triangle} and \textit{square} mark selected positions where ambient conditions are estimated from. 
The reference position ($\Delta\alpha=0$, $\Delta\delta=0$), marked with a cross, corresponds to the SAO star 161357 at R.A(J2000)=18:20:27.6483 and Dec(J2000)=-16:12:00.9077.}}

  \label{fig:set1-map}
\end{figure}

\begin{figure}[!ht]
  
  \hfill\includegraphics[width=6.5cm,angle=-90]{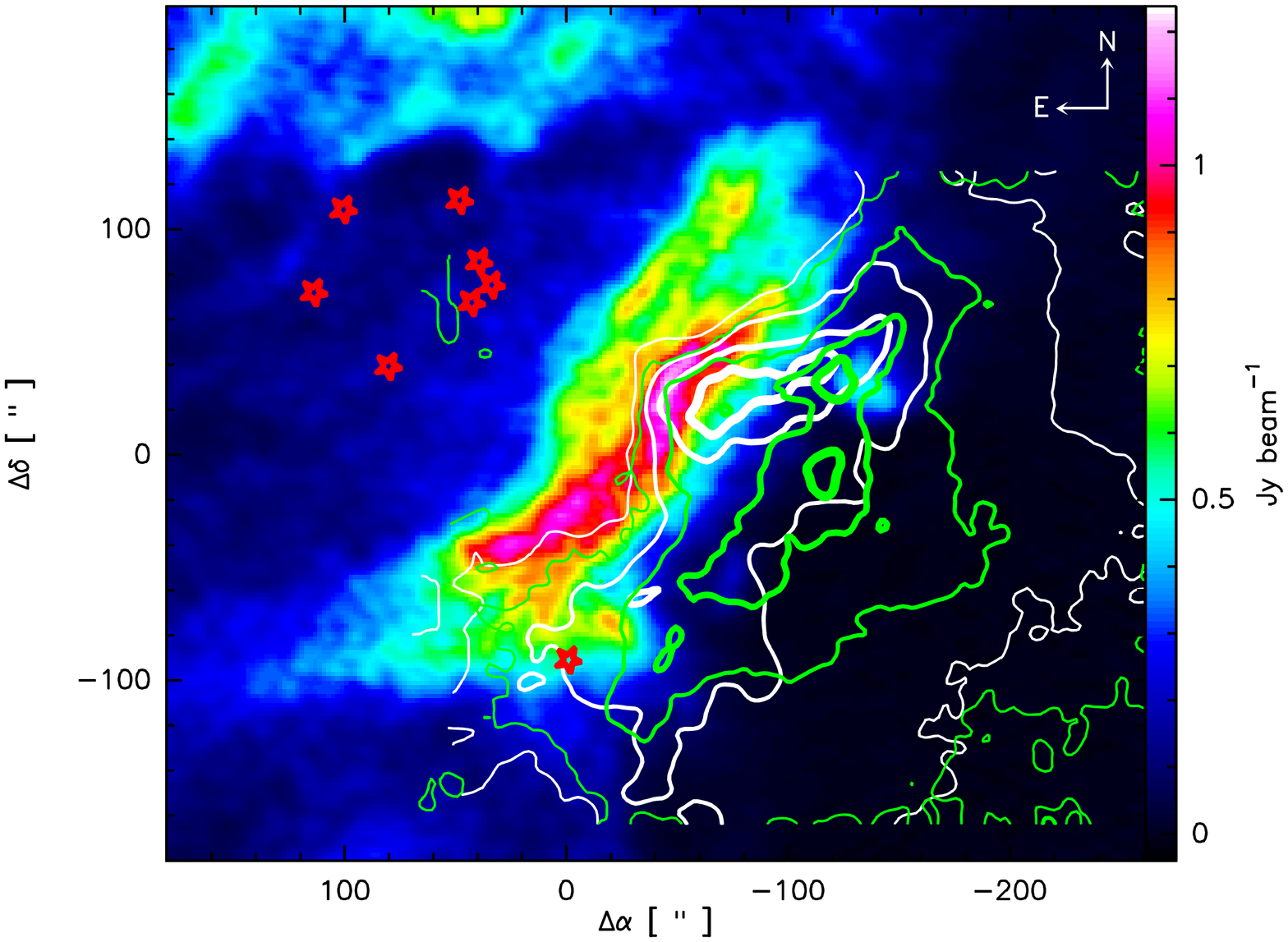}\hspace*{\fill}

  \caption{\footnotesize{Color map of the 21 cm continuum emission (Jy beam$^{-1}$) in M17 SW with $10''\times7''$ resolution by Brogan \& Troland (2001). The \textit{white} contour lines correspond to the \twco $J = 6\rightarrow5$ with a peak emissions of 830 K \kms. While the \textit{green} contour lines correspond to the \ci $J = 2\rightarrow1$ with a peak emissions of 260 K \kms. The countour levels (from thin to thick) are the 25\%, 50\%, 75\% and 90\% of the peak emission. The \textit{red stars} indicates the O and B ionizing stars (Beetz \etal\ 1976; Hanson \etal\ 1997). The reference position ($\Delta\alpha=0$, $\Delta\delta=0$) is the same as in Fig~\ref{fig:set1-map}. These \twco and \ci maps have a slightly lower integrated temperature than in Fig.~\ref{fig:set1-map} because they were convolved with a $20''$ beam to smooth the contour lines.}}

  \label{fig:vla-map}
\end{figure}

We used the Fast Fourier Transform Spectrometer (FFTS) as backend with a fixed bandwidth of 1.5 GHz and
1024 channels. We used the two IF groups of the FFTS with an offset of $\pm460$ MHz between them.
The spectral resolution was smoothed to about 1 \kms, while the line widths are between 4 \kms\ and 9 \kms, so they are well resolved.
The on source integration time per dump and pixel was 1 second only.
However, oversampling with $4''$ spacing, all the seven pixels of CHAMP$^+$ covered a given grid position 
at least once. So, after adding all the subscans from both IF channels, and after convolving the maps with the corresponding beam size, the total integration time in the central 5'x4' region of the maps varies between about 50 and 80 seconds per grid cell.

The SSB system temperatures are typically about 2000 K and 6000 K for the low and high frequency 
bands, respectively. The spatial resolution varies between $9.4''$ for \thco $J = 6\rightarrow5$ transition in the low 
frequency band (at 661 GHz - the nominal beam at 691 GHz is $8.4''$) and $7.7''$ for the high frequency band (809 GHz).
All data in the paper were converted to line brightness temperature, $T_{B}=\eta_{f}\times T_{A}^{*}/\eta_{c}$, using a forward efficiency ($\eta_f$) of 0.95 and beam coupling efficiencies ($\eta_c$) of 0.45 and 0.43 (at 661 GHz and 809 GHz, respectively) as determined towards Jupiter\footnote{http://www.mpifr.de/div/submmtech/heterodyne/champplus/ champmain.html} (G\"usten \etal\ 2008). We assumed brightness temperatures of 150 K (at 660 GHz) and 145 (at 815 GHz) for Jupiter (Griffin \etal\ 1986).
This coupling efficiency was chosen because in velocity-space (velocity channels) the size of the M17 clumps is Jupiter-like, which had a size $\sim38.7''$ by the time of the observations.
The calibrated data were reduced with the GILDAS\footnote{http://www.iram.fr/IRAMFR/GILDAS} package CLASS90.

\section{Results}

\subsection{Integrated line temperature maps}

\begin{figure}[!ht]
  
  \hfill\includegraphics[width=7cm, angle=-90]{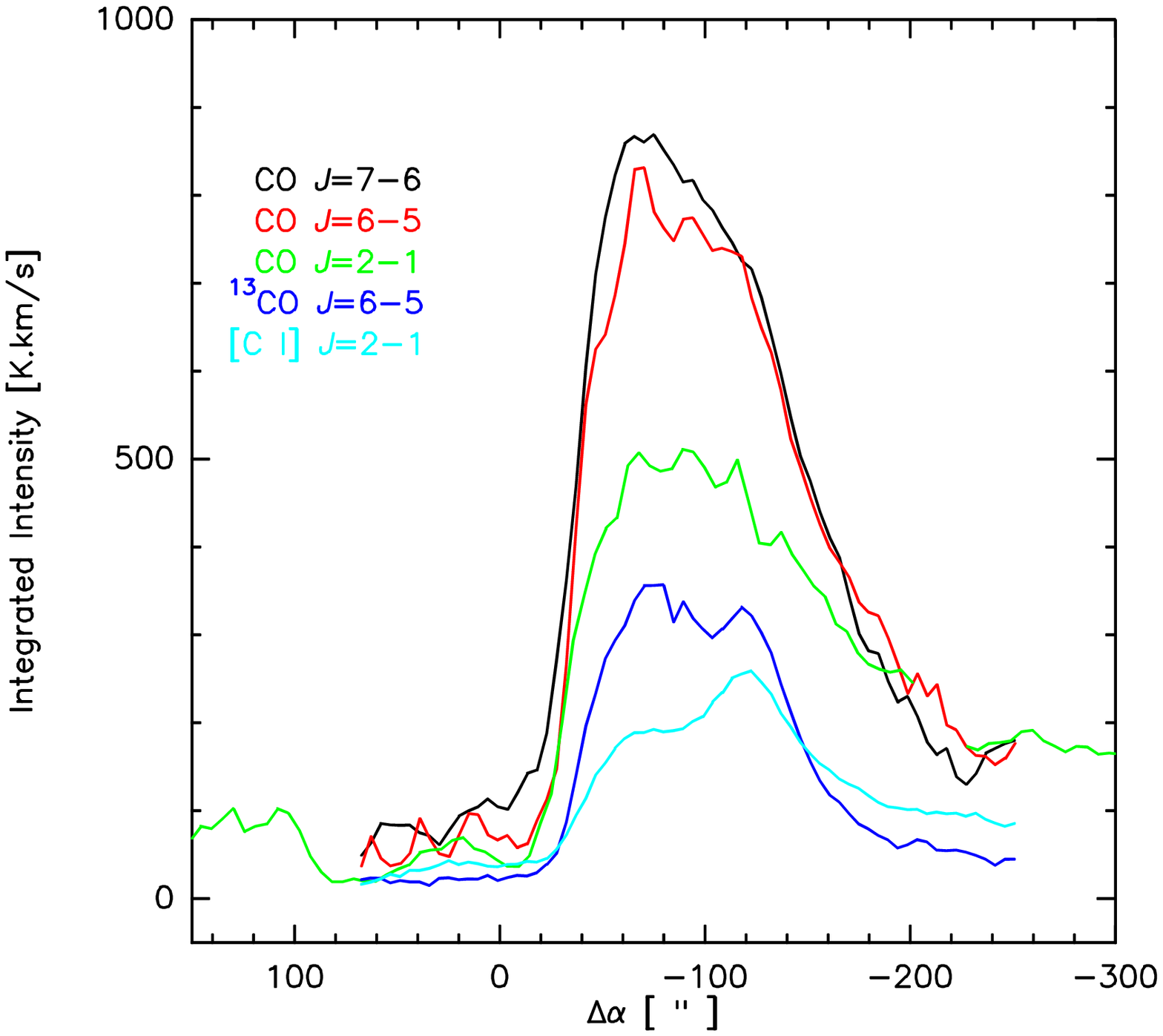}\hspace*{\fill}\\

  \hfill\includegraphics[width=7cm, angle=-90]{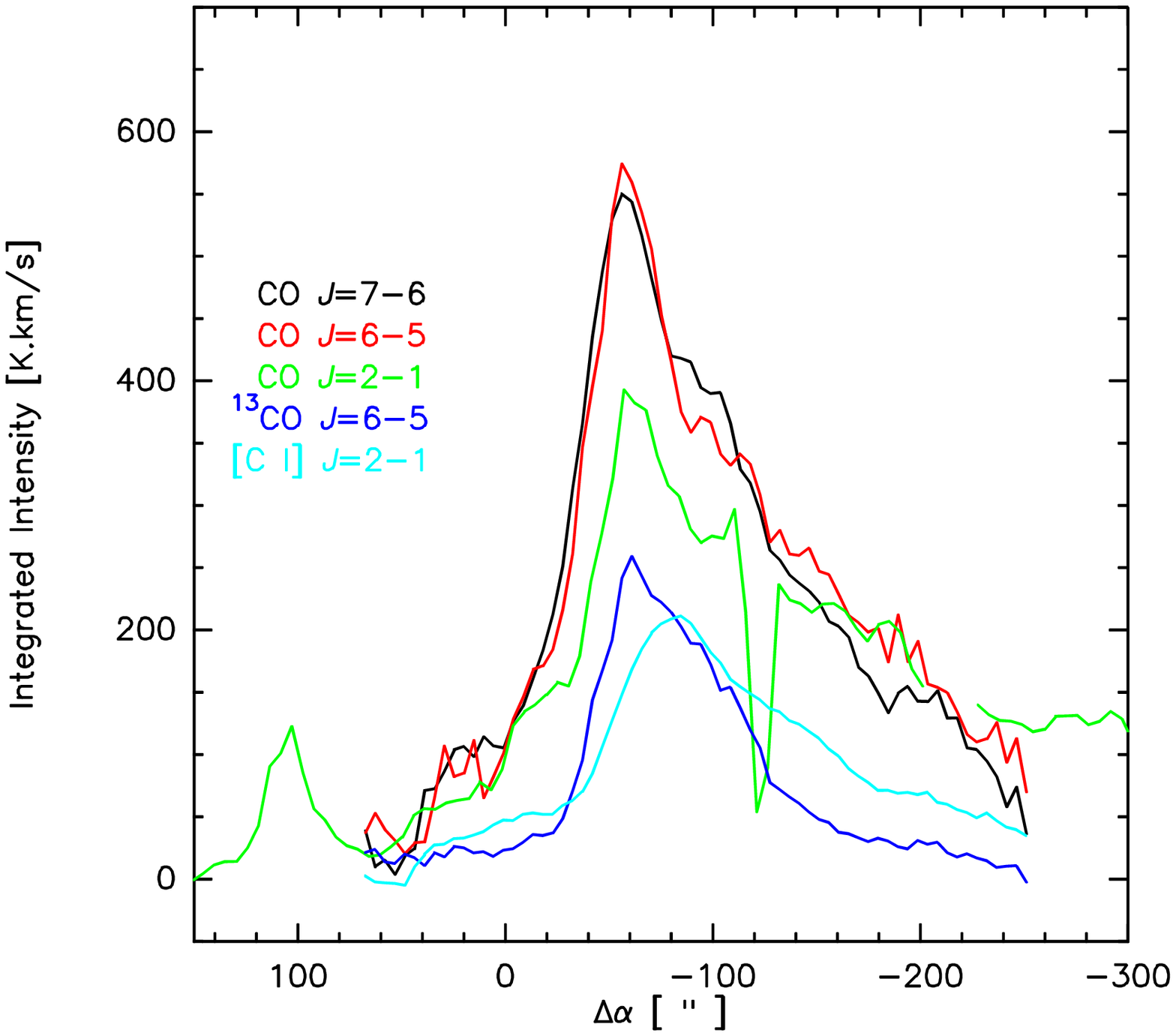}\hspace*{\fill}\\
 
  \caption{\footnotesize{\textit{Top panel} - Strip lines of the velocity integrated intensities of \twco $J$=7--6 (black), \twco $J$=6--5 (red), \twco $J$=2--1 (green) (adapted from S88), \thco\ $J$=6--5 (blue) and \ci\ $J$=2--1 (cyan) at $\Delta\delta=30''$ (P.A. $90^{\circ}$) across the ionization front of M17 SW. \textit{Bottom panel} - Strip lines at P.A. $63^{\circ}$ ($\Delta\delta=\Delta\alpha/2$). The X-axis corresponds to the actual offset in R.A. of the maps shown in Figure~\ref{fig:set1-map}. Hence, $\Delta\alpha=0''$ is the R.A. of the reference illuminating star SAO 161357.}}


  \label{fig:strip-lines}
\end{figure}

Figure \ref{fig:set1-map} shows the maps of the temperature, integrated between 5 \kms\ and 35 \kms, of
\twco $J = 6\rightarrow5$ (\textit{top}) with the contour lines of \twco $J = 7\rightarrow6$, 
and the velocity integrated temperature of \thco $J = 6\rightarrow5$ (\textit{bottom}) with the 
contour lines corresponding to \ci $J = 2\rightarrow1$.
All the maps were convolved to the largest beam size ($9.4''$) of the
\thco $J = 6\rightarrow5$ line, obtaining a grid size of about $4.7''\times4.7''$. The peak
integrated temperature of the \twco $J = 6\rightarrow5$ and $J = 7\rightarrow6$ lines are 852 K \kms\ and 925 K \kms\,
respectively. These lines follow a similar spatial distribution.
The peak integrated temperature of \thco $J = 6\rightarrow5$ and \ci $J = 2\rightarrow1$ are 420 K and 282 K.
respectively, and the peak of \ci is shifted towards the inner side of the interface region at about 0.55 pc ($\sim50''$).
The ionization front traced by the high resolution ($10''\times7''$) map of the 21 cm continuum emission (Brogan \& Troland 2001), as well as the ionizing stars identified by Beetz \etal\ (1976) and Hanson \etal\ (1997), are shown in Figure~\ref{fig:vla-map}, with \twco $J = 6\rightarrow5$ (white contour lines) and \ci $J = 2\rightarrow1$ (green contour lines) overlaid. The transition between the hot ($T_k>300$ K) atomic gas and the warm ($T_k>100$ K) molecular gas can be seen thanks to the almost edge-on geometry of M17 SW.

The \textit{top panel} of Figure~\ref{fig:strip-lines} shows the variation of the integrated temperature of all the lines, across the ionization front (strip line at P.A=$90^{\circ}$ in Fig.~\ref{fig:set1-map}). {Due to the limited S/N the \twco $J = 7\rightarrow6$ and \ci $J = 2\rightarrow1$ strip lines have been smoothed spatially with respect to the strip direction.
The \ci\ $J=2\rightarrow1$ line starts peaking up at about 0.1 pc ($\sim10''$) after the molecular lines and presents a smooth transition towards the inner part of the cloud, forming a plateau up to about $\Delta\alpha=-100''$, from where it increases its emission until the peak reached at about $\Delta\alpha=-120''$. The peak of \ci\ correlates with a secondary peak seen in \thco. However, the main peak emission of the latter correlates with the peak of the \twco\ lines along this strip line.

The strip line at P.A.=$63^{\circ}$ (\textit{bottom panel} of Figure~\ref{fig:strip-lines}) can be compared with Fig.5 in M92, and Fig.2 in S88. At this position angle, there is not a marked plateau in the \ci\ emission, and the peak of the \ci\ line is closer to the peaks of the \twco\ and \thco\ lines. 
The dip in \twco $J = 2\rightarrow1$ at about $\Delta\alpha=-150''$ is an artifact.
The integrated temperature of the \twco $J=2\rightarrow1$, $J=6\rightarrow5$ and $J=7\rightarrow6$ lines have comparable strength deep ($\Delta\alpha>-160''$) into the M17 SW complex,



\subsection{The complex internal structure of M17 SW}

Figure~\ref{fig:strip-lines2} shows the spectra at selected positions along the NE-SW strip line at P.A. $63^{\circ}$. The main beam temperature of the spectra is shifted 70 K at each offset position.
This set of spectra can be compared with the \twco\ and C$^{18}$O $J=2\rightarrow1$ spectra along the same strip line of Fig.8 in S88. The warm gas ($T_K>50$ K), traced by the mid-$J$ \twco\ lines, is as extended as the cold gas ($T_K<50$ K) traced by the \twco $J=2\rightarrow1$ line deeper into the cloud. On the other hand, the \thco\ $J=6\rightarrow5$ and \ci\ $J=2\rightarrow1$ lines are strongly detected in a narrower spatial extent of about 1.3 pc, similar to the extent of the C$^{18}$O $J=2\rightarrow1$ emission.




\begin{figure*}[!ht]
  
  \hfill\includegraphics[height=14cm,angle=0]{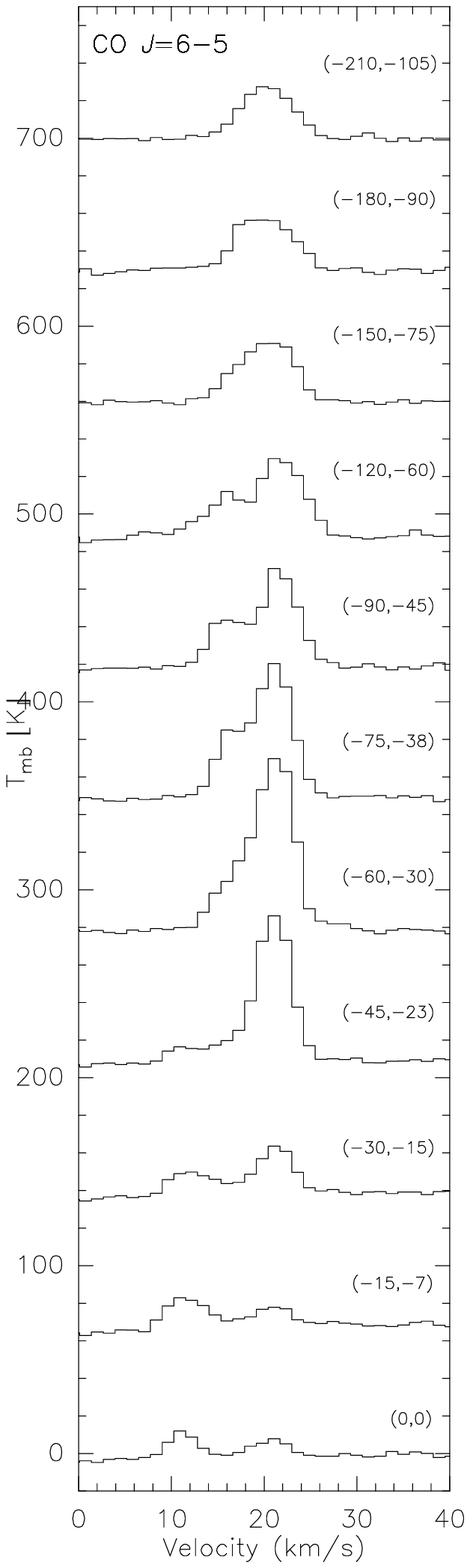}\hspace{-0.2cm}
  \includegraphics[height=14cm,angle=0]{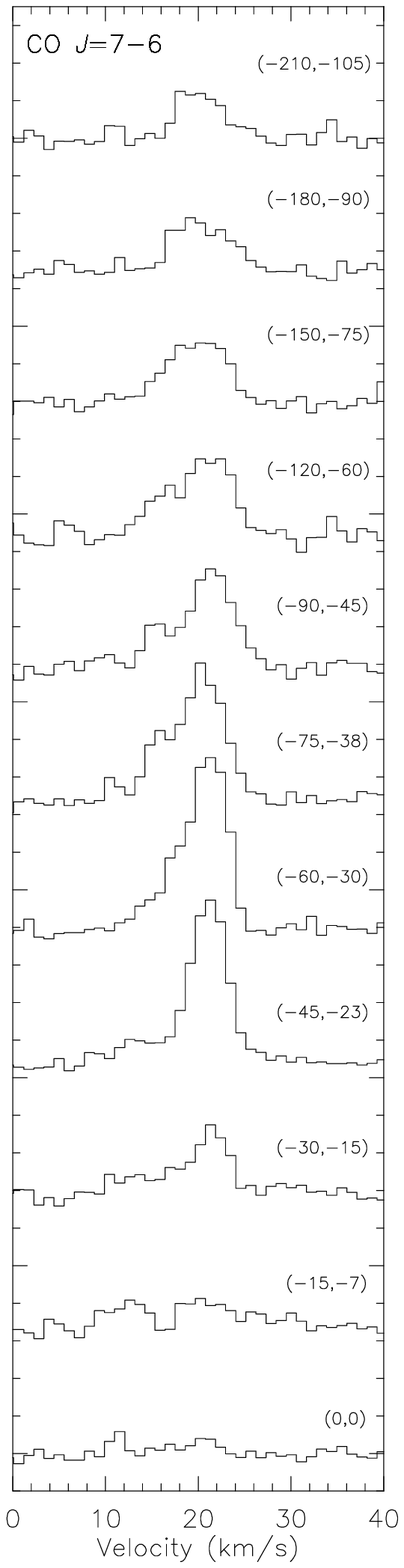}\hspace{-0.2cm}
  \includegraphics[height=14cm,angle=0]{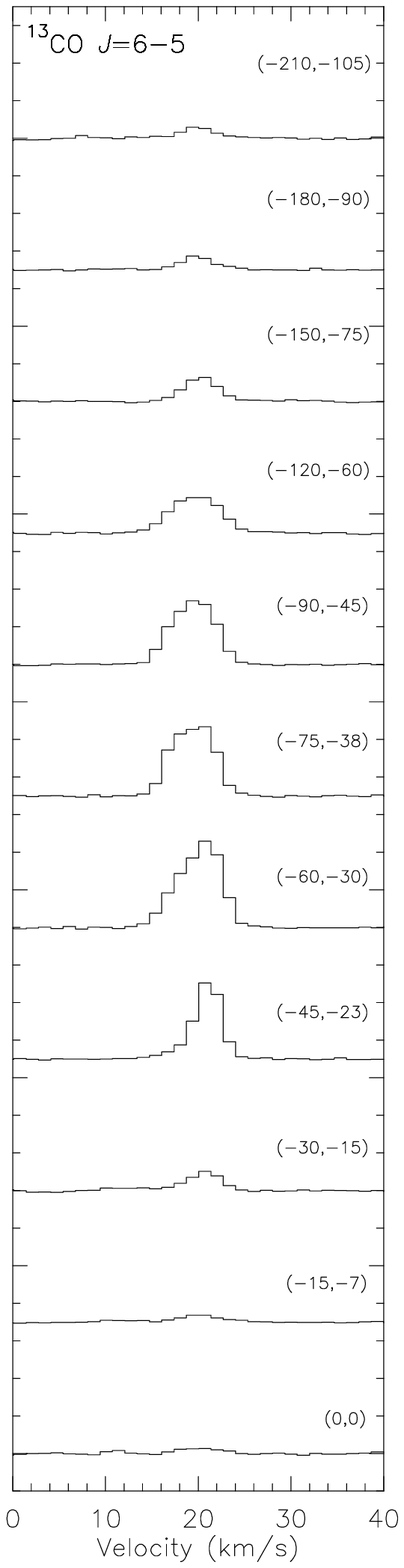}\hspace{-0.2cm}
  \includegraphics[height=14cm,angle=0]{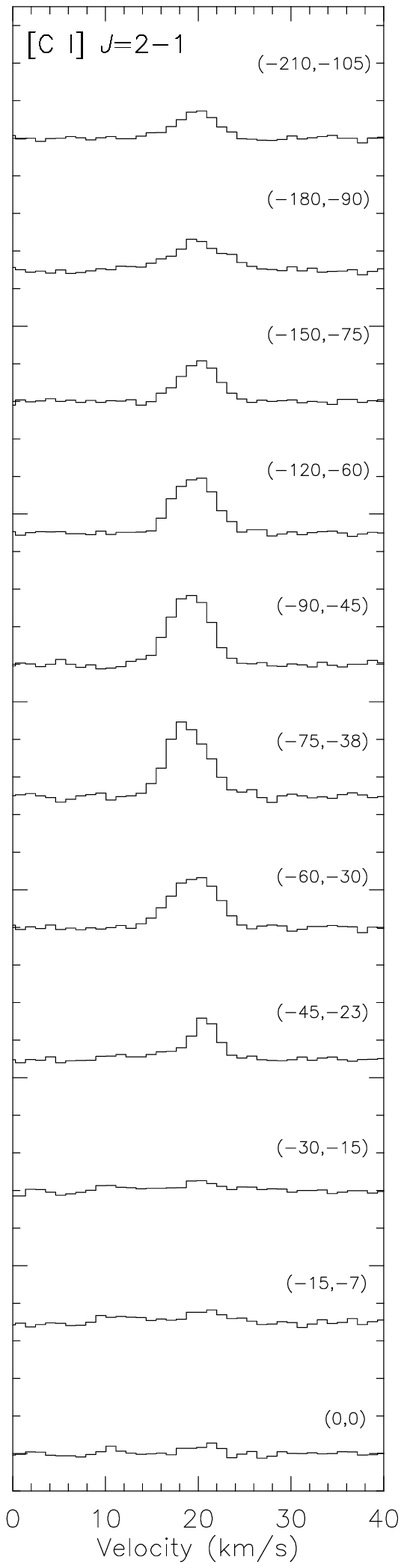}\hspace*{\fill}\\

  \caption{\footnotesize{Selected spectra of \twco\ $J = 6\rightarrow5$, \twco\ $J = 7\rightarrow6$, \thco\ $J = 6\rightarrow5$ and \ci\ $J = 2\rightarrow1$ along the NE-SW strip line (see Fig.~\ref{fig:set1-map}). The position in arcsecs are the offsets with respect to the reference position at R.A(J2000)=18:20:27.6483 and Dec(J2000)=-16:12:00.9077. The spectra are the average spectra within $\pm2''$ of the indicated offset positions, and smoothed to 1 \kms\ velocity resolution.}}

  \label{fig:strip-lines2}
\end{figure*}

Multilevel molecular line observations in CS, \twco, \thco and C$^{18}$O, and in several fine structure lines (\ci, \cii, [Si II],[O I]) indicate that M17 SW consists of numerous high density clumps ($n(\rm H_2)>10^4~\3cm$) where the [O I], [Si II] and mid-$J$ CO lines emanate from. This dense gas is found within relatively warm ($\sim50$ K) and less dense ($n(\rm H_2)\sim3\times10^3~\3cm$) molecular gas (interclump medium) which in turn is surrounded by a diffuse halo ($n(\rm H_2)\sim300~\3cm$) where the very extended  \ci and \cii emission emerge from (Snell \etal\ 1984, 1986; Evans \etal~1987; S88; SG90; M92).


From \c18o observations in M17 SW a beam-averaged ($13''$) column density of $\sim8\times10^{23}~\2cm$ has been estimated for the cloud core, and masses in the range $\sim10-2000~M_{\sun}$ for the CO clumps (SG90). A comparable mass range ($\sim10-120~M_{\sun}$) was lately estimated from submillimeter continuum observations in the northern part of M17 (Reid \& Wilson 2006). Although the region mapped by Reid \& Wilson (2006) adjoins, but does not overlap with M17 SW.

Figure~\ref{fig:set3-channel} shows representative velocity channel maps of the \twco $J = 6\rightarrow5$ (\textit{top left}) and $J = 7\rightarrow6$ (\textit{top right}) lines in M17 SW. These are the main-beam brightness temperature averaged over 2 and 3 velocity channels between 18.2 \kms\ and 19.9 \kms. These are similar velocity channels shown in Fig.3 by SG90. The fact that the C$^{18}$O $J=2\rightarrow1$ line trace colder ($T_K<50$ K) and less dense ($n_H\sim3\times10^3~\3cm$) gas than the \twco lines is reflected in the different velocity integrated and channel maps of these lines. In theory the critical densities (at $T_K=100$ K) of the \twco $J=6\rightarrow5$ and $J=7\rightarrow6$ lines are $n_{crit}\sim2.7\times10^5~\3cm$ and $n_{crit}\sim4.4\times10^5~\3cm$, respectively, which corresponds to a factor $\sim1.6$ difference. However, this difference does not translates directly into different clumpyness. This is reflected in the similar clumpy structure seen in the channel maps of these mid-$J$ \twco lines.


Even though the critical density of the \thco $J = 6\rightarrow5$ line is similar to that of the \twco ($n_{crit}\sim2.4\times10^5~\3cm$) the South-East region of its channel map (\textit{bottom left}) differs from that seen with the \twco lines. This could be due to a change in the temperature of the gas, or to a variation in the \thco column density in that region. Since \thco is much more optically thin than \twco (abundance ratio of about $50-70$) this difference in the map can be expected. In Sections 4.2 and 4.3 we discuss about the optical depths.

On the other hand, the \ci $J = 2\rightarrow1$ channel map (\textit{bottom right}) shows a completely different structure and distribution than the \twco and the isotope lines. Since the critical density of this line is about $2.8\times10^3~\3cm$, its emission is likely emerging partly from the interclump medium mentioned above.

\begin{figure}[!ht]
  
  \hfill\includegraphics[width=3.3cm,angle=-90]{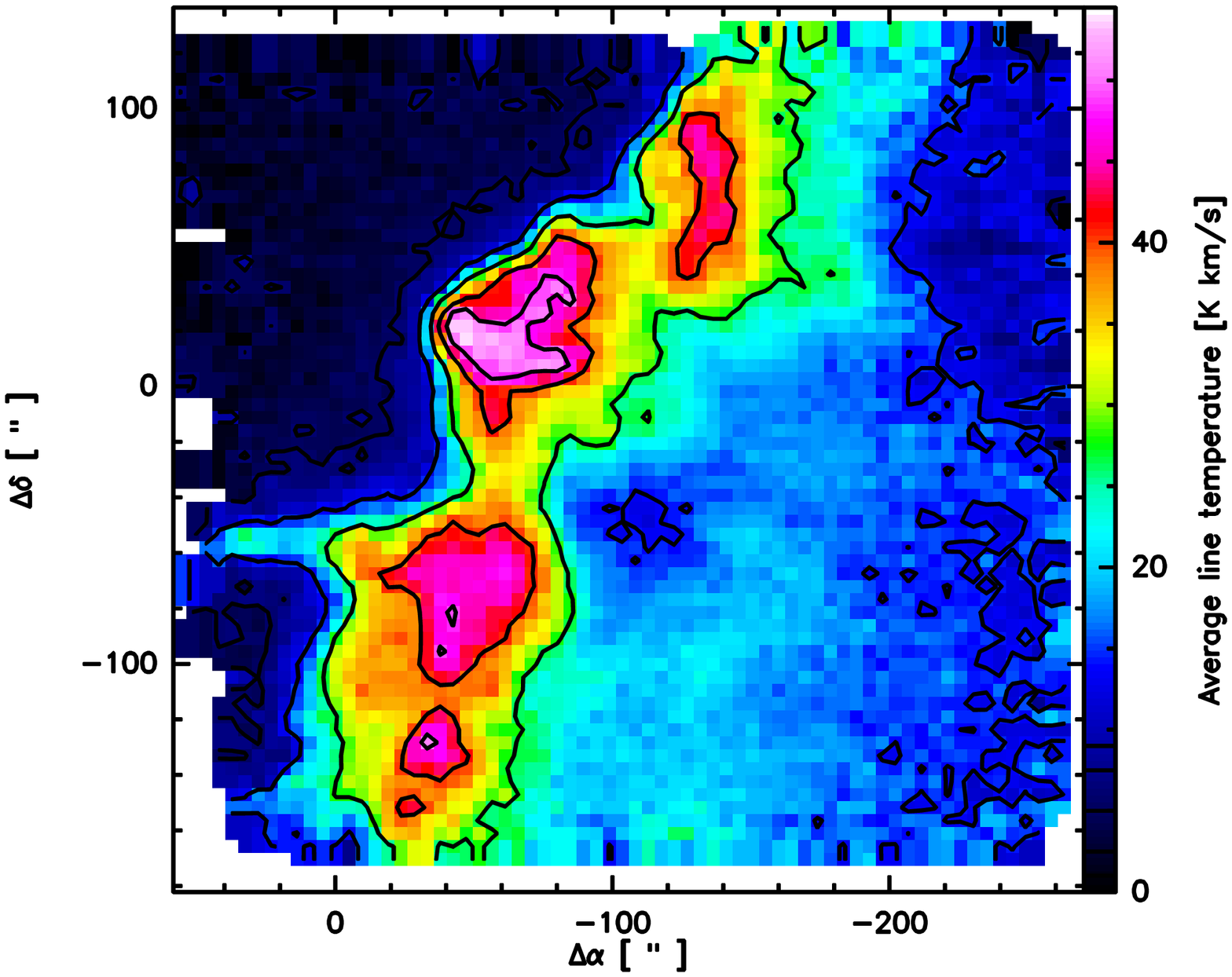}
  \hspace*{\fill}\includegraphics[width=3.3cm,angle=-90]{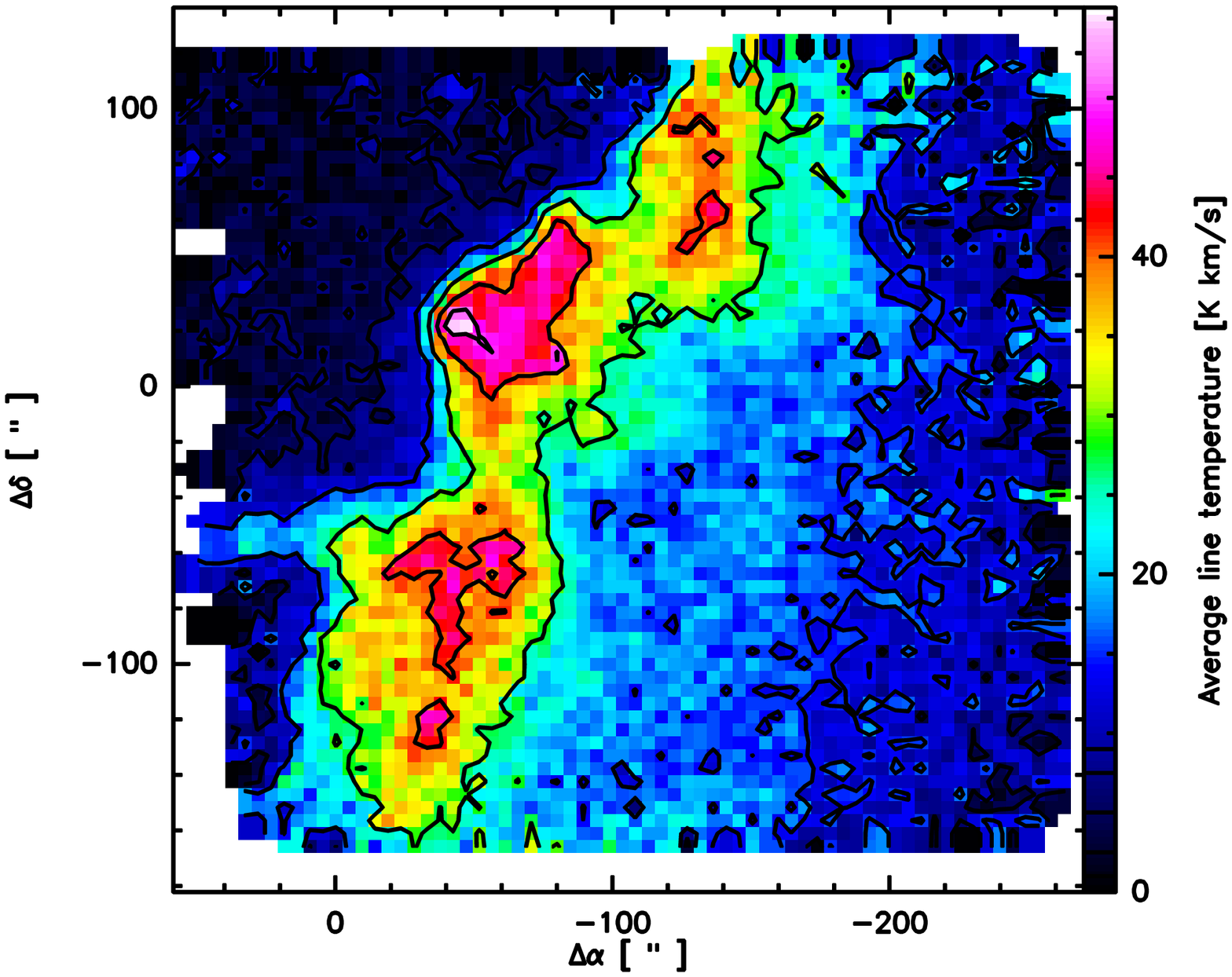}\hspace*{\fill}\\

  \hfill\includegraphics[width=3.3cm,angle=-90]{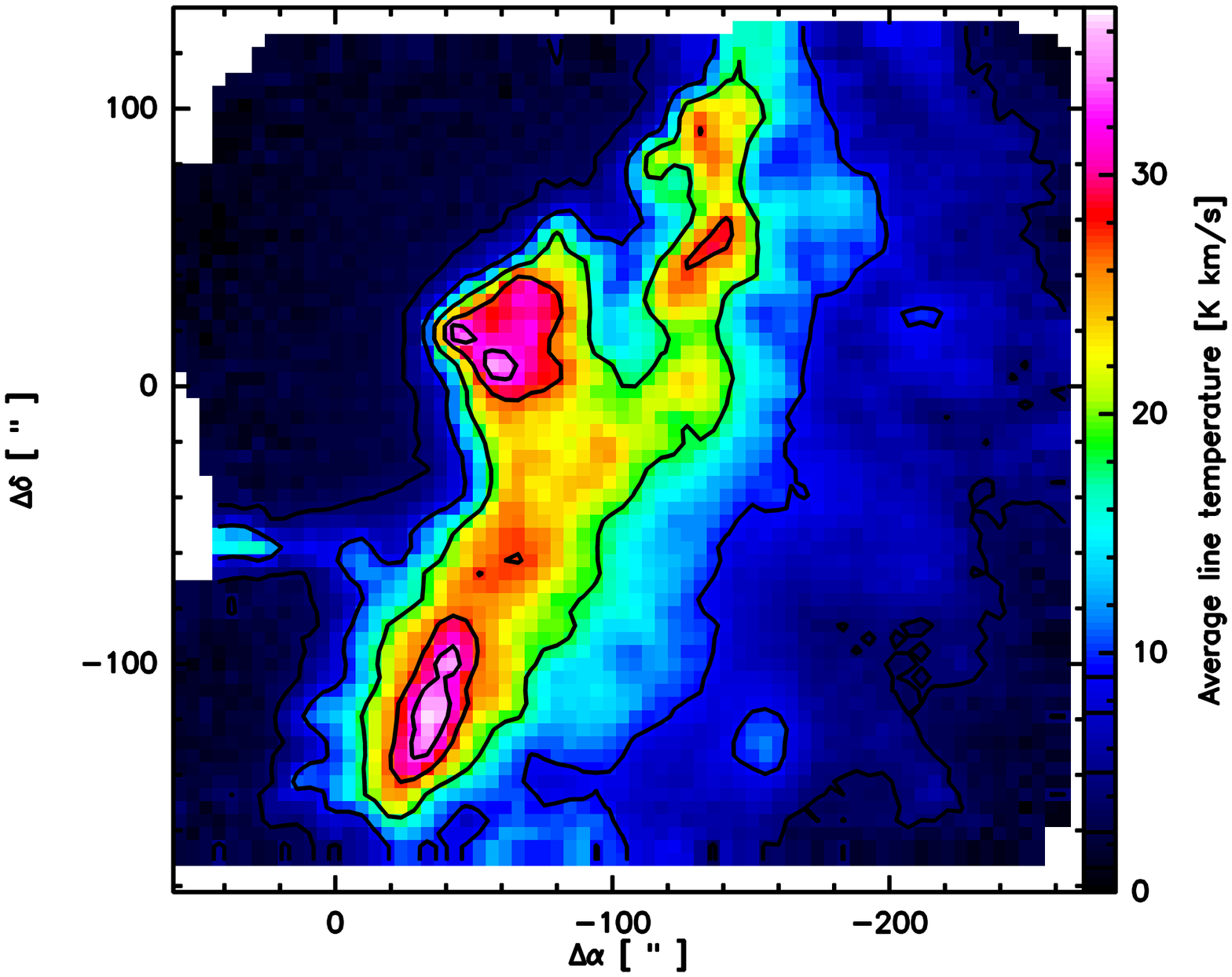}
  \hspace*{\fill}\includegraphics[width=3.3cm,angle=-90]{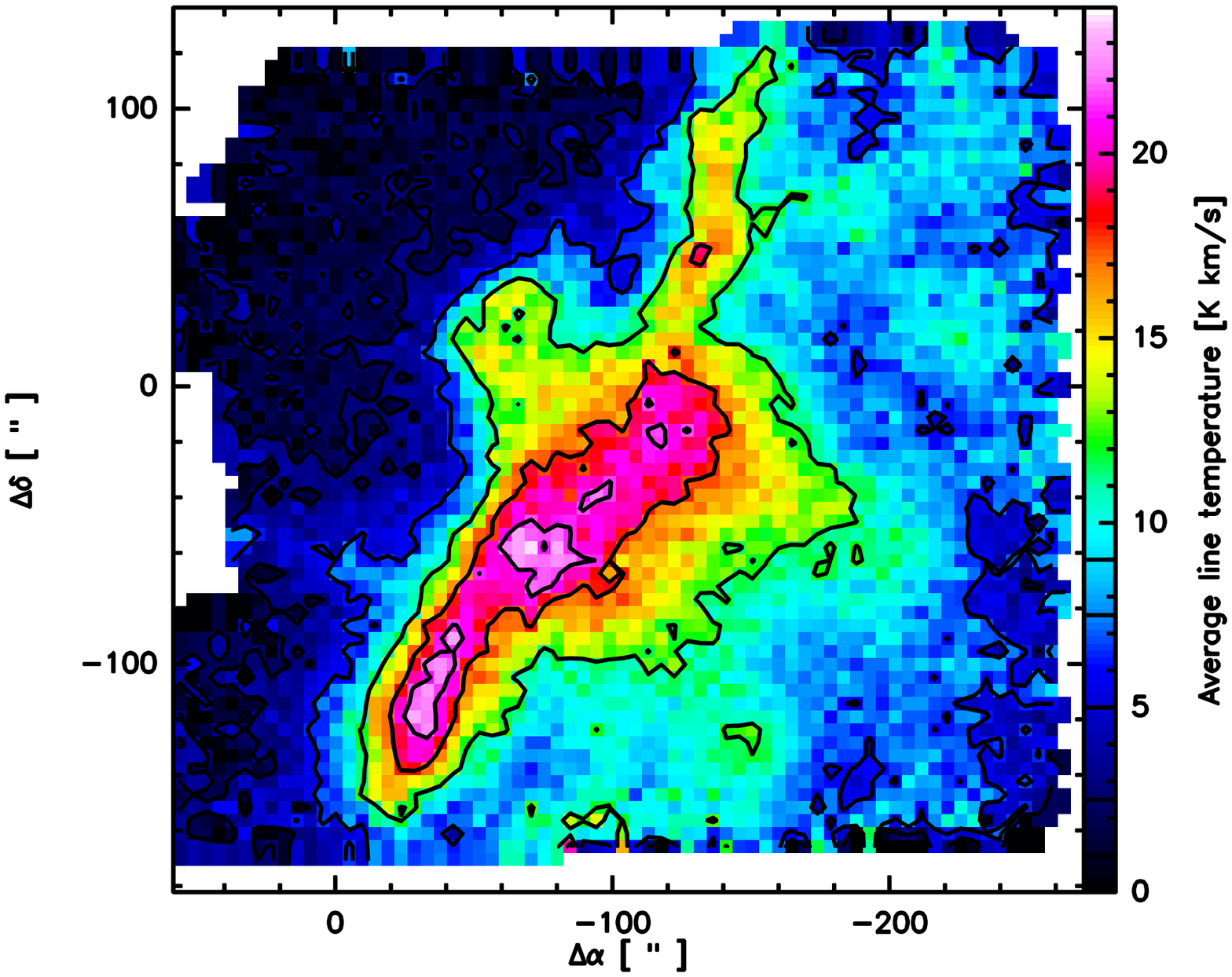}\hspace*{\fill}\\
  

 
  \caption{\footnotesize{\textit{Top panel} - Channel maps of the main-beam brightness temperature of \twco $J = 6\rightarrow5$ (\textit{left}) and $J = 7\rightarrow6$ (\textit{right}) averaged over 2 and 3 velocity channels, respectively, between 18.2--19.9 \kms. The contour lines are as described in Fig.1, with peak integrated line temperatures of 54.5 \Kkms and 55.7 \Kkms for the $J = 6\rightarrow5$ and $J = 7\rightarrow6$ lines, respectively.
  \textit{Bottom panel} - Channel maps of \thco $J = 6\rightarrow5$ (\textit{left}) and \ci $J = 2\rightarrow1$ (\textit{right}),
  averaged over 2 velocity channels between 18.4--19.7 \kms. Contours are as in top, and the peak integrated line temperatures are 37.0 \Kkms and 23.9 \Kkms.
}}

  \label{fig:set3-channel}
\end{figure}

\section{Discussion}

\subsection{Self-absorption in the mid-$J$ \twco lines?}

The complex structure of the \twco $J = 1\rightarrow0$, $J = 2\rightarrow1$, and $J = 3\rightarrow2$ 
line profiles has been attributed to strong self-absorption effects (e.g. Rainey \etal\ 1987; Stutzki \etal 1988). 
Martin, Sanders \& Hills (1984) also reported a flat topped spectra of \twco $J = 3\rightarrow2$, 
attributed to self-absorption or saturation at velocities near the line center, and give details about 
the effects of macroturbulent clumpy medium in line profiles.

A double peaked structure in the \thco $J = 1\rightarrow0$ line was also reported by Lada (1976). 
Rainey \etal\ considered that this double peaked structure in \thco suggest that either this line 
is optically thick, or that the double peaked structure is due to more than one cloud component. 
The latter is the interpretation favored by Rainey \etal\ in view of the available data at that time. 

Phillips \etal\ (1981) presented a self-absorption LTE model that considers a \twco\ cloud of uniform  
temperature $T_k$ in front of a hot background source of temperature $T_{bg}$, at the same central velocity. 
The velocity dispersion of the background cloud is considered to be larger compared to that of the foreground cloud,
so the self-absorption effect is seen mostly at the line center.
This model indicates that, depending on the total column density of \twco, the self-absorption effect will be stronger in 
the $J = 2\rightarrow1$ and $J = 3\rightarrow2$ lines than in the $J = 1\rightarrow0$ line, 
with decreasing intensity as the transition number $J$ increases. This is indeed observed in Fig.12 of S88,
for the \twco $J = 4\rightarrow3$, $J = 3\rightarrow2$, $J = 2\rightarrow1$, and $J = 1\rightarrow0$ lines.

We reproduced the model by Phillips \etal\, including the higher-$J$ lines of \twco.
The \textit{top panel} of Figure~\ref{fig:self-abs-models} shows the model with the same background and foreground temperatures used by Phillips \etal. This model implies that, for a background temperature $T_{bg}=64$ K and a foreground kinetic temperature $T_k=15$ K, the lower-$J$ lines (J=1,2,3,4) of the background cloud start showing self absorption at the line center for lower column densities ($N/\Delta V$ = $10^{14}$ - $10^{15}$ cm$^{-2}$ km s$^{-1}$). Instead, the higher-$J$ lines 
(J=5,6,7) need larger columns ($N/\Delta V$ = $10^{15}$ - $10^{17}$ cm$^{-2}$ km s$^{-1}$) in order to be affected by 
self-absorption. For a velocity dispersion $\Delta V=5$~\kms, the upper limits of these \twco\ columns would correspond to extinctions $A_v$ of $\sim0.1$ mag and $\sim10$ mag, respectively.

The \textit{bottom panel} of of Figure~\ref{fig:self-abs-models} shows the model for a background temperature $T_{bg}=150$ K and a foreground temperature $T_k=30$ K (from S88). In this case the lower-$J$ lines show self-absorption at the same range of columns as before, while the higher-$J$ lines start showing self-absorption at a narrower range of columns ($N/\Delta V$=$10^{15}$ - $10^{16}$ cm$^{-2}$ km s$^{-1}$). A remarkable characteristic of these 
models (top and bottom panels of Figure~\ref{fig:self-abs-models}) is that all the $J$ lines are expected to be strongly self-absorbed at columns larger than $10^{18}$ cm$^{-2}$ km s$^{-1}$, which is similar to the column density estimated by S88. Another characteristic is that the \twco emission of the higher-$J$ lines are also expected to decrease with the transition number $J$, and be weaker than the low-$J$ lines. However, the \twco $J = 7\rightarrow6$ line seems to break this rule, as can be seen in fig.12 of S88. The high peak temperature observed in the \twco $J = 7\rightarrow6$ line is missing in the lower-$J$ lines. Even considering a calibration uncertainty of 20\%, the \twco $J = 7\rightarrow6$ line (observed at offset position $(-100'',0'')$, bottom panel of Fig.12 in S88) will be as strong as the $J = 4\rightarrow3$ line (at 
least at the peak intensity) but still stronger than the $J = 2\rightarrow1$ line. 

\begin{figure}[!t]
  
  \hfill\includegraphics[width=7cm, angle=0]{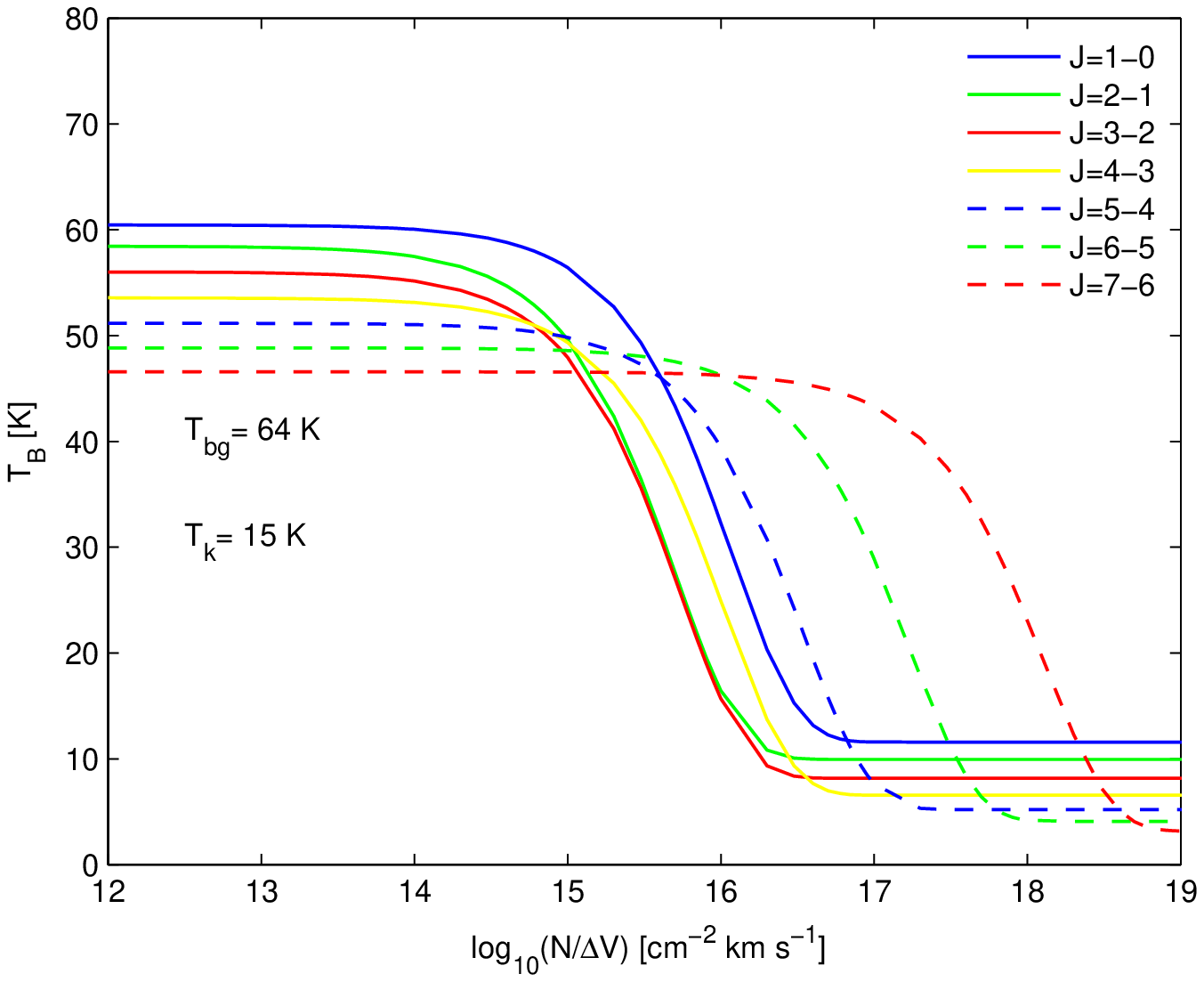}\hspace*{\fill}\\

  \vspace{-0.4cm}
  \hfill\includegraphics[width=7cm, angle=0]{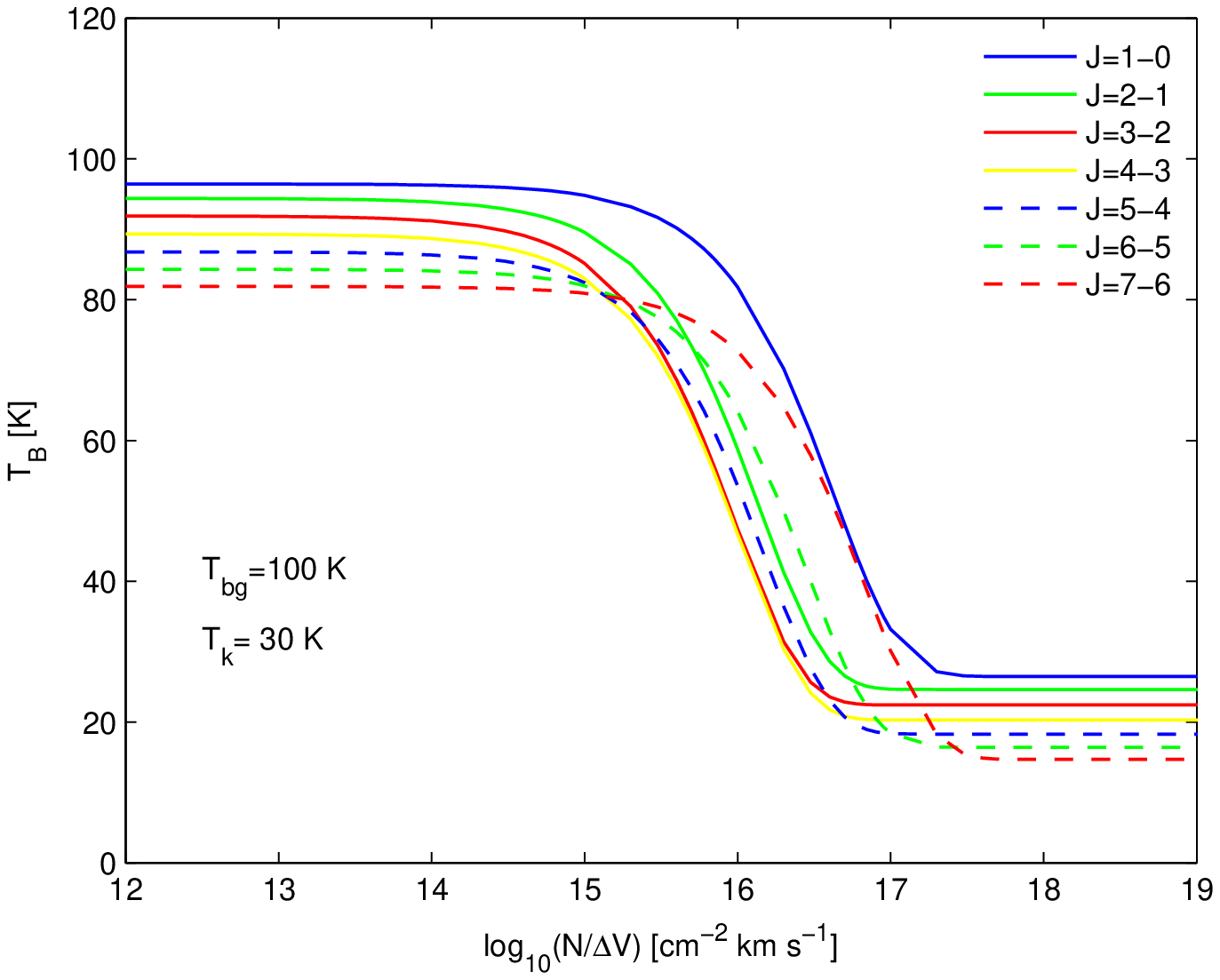}\hspace*{\fill}\\
  \vspace{-0.3cm}
 
  \caption{\footnotesize{\textit{Top panel} - Expected brightness temperature at the center of the \twco\ lines for a warm background cloud with temperature $T_{bg}=64$ K and a colder foreground absorbing cloud with temperature $T_k=15$ K. \textit{Bottom panel} - Same as in top, but for a background temperature $T_{bg}=100$ K and a foreground temperature of $T_k=30$ K.}}

  \label{fig:self-abs-models}
\end{figure}

On the other hand, the \twco $J = 7\rightarrow6$ line seems to be asymmetric, with a \textit{left shoulder} weaker than the
\textit{right shoulder}, which may be due to self-absorption produced by a colder foreground cloud with 
slightly lower center velocity than the warmer clump traced by the $J = 7\rightarrow6$ line. However, that 
weaker left shoulder of the mid-$J$ line is still brighter than the corresponding shoulder of the
lower-$J$ lines, in most of the velocity range and in both positions $(-100'',0'')$ and $(-60'',-30'')$ - assuming low ($<10$\%) uncertainty in the calibration of the data. This is not what would be expected in the self-absorption scenario proposed by Phillips \etal\ (1981).

Figure~\ref{fig:strip-lines2} shows that the \thco $J = 6\rightarrow5$ line has a similar 
asymmetry as the \twco lines, at positions $(-45'',-23'')$ and $(-60'',-30'')$. But it shows only one component
at the other positions. This difference may be related to a gradient in the temperature (or total column density) of the foreground cloud that produces self-absorption in the first two positions, but not in the others.
Instead, \ci\ $J = 2\rightarrow1$ shows similar asymmetry as \twco\ at positions $(-45'',-23'')$, $(-150'',-75'')$ and $(-180'',-90'')$, and an opposite asymmetry at position $(-75'',-38'')$. Given that there is no strong evidence for self-absorption in the \thco\ lines, nor in the \ci\ lines, and that the \thco lines are mostly optically thin, it is unlikely that the observed asymmetries of the \thco and \ci lines are produced by self-absorption. Hence, we agree with Rainey \etal\ (1987) in that this complex structure is more likely due to more than one kinematical component along the line of sight. And this could also be the case for the mid-$J$ \twco lines.

Therefore, the observational facts and the models suggest that the self absorption effect, if present, should have little impact on the mid-$J$ lines, and a few cloud components at different central velocities could also explain the complex structure of the line profiles. The asymmetry of the profiles suggests that self-absorption affects mostly one wing of the line profile, while the peak temperatures seems to be the least affected velocity channel in the mid-$J$ lines.
Hence, in the following sections we test the ambient conditions of the warm gas based on the ratios between the peak main-beam temperatures of the \twco and \thco $J = 6\rightarrow5$ and $J = 7\rightarrow6$ lines.

\subsection{Optical depth and excitation temperature (LTE)}

Since we have the maps of \twco and \thco $J=6\rightarrow5$ lines, we can estimate the optical depth and the excitation temperature of these lines, assuming local thermal equilibrium (LTE), from the ratio between their peak main beam temperature $T_{mb}$ observed between 5 \kms\ and 35 \kms velocity channels. This will provide at least a lower limit for the kinetic temperature in M17 SW. Then we will estimate the ambient conditions at two selected positions based on a non-LTE model of the ratio between the peak $T_{mb}$ temperatures of the \twco $J=6\rightarrow5$ and $J=7\rightarrow6$ lines (hereafter referred as \twco $\frac{7-6}{6-5}$ line ratio). The temperature and densities obtained in this way will be compared to those values estimated in previous work. 

In LTE the radiation temperature can be approximated (e.g. Kutner 1984; Bergin \etal\ 1994) by the expression:

\begin{equation}
 T_{R}=[J_{\nu}(T_{ex})-J_{\rm bg}][1-e^{-\tau_{\nu}}] ,
\end{equation}

\noindent
where the term $J_{\nu}(T)$ is the Planck's function evaluated at
frequency $\nu$ and temperature $T$, and multiplied by the factor $\frac{\lambda^2}{2k}$ to obtain the intensity in K. So it is defined as:

\begin{equation}
 J_{\nu}(T)=\frac{\nicefrac{h\nu}{k}}{e^{\nicefrac{h\nu}{kT}}-1}.
\end{equation}


We use the full $J_{\nu}(T)$ function since the Rayleigh-Jeans (R-J) approximation 
(commonly applied when $h\nu\ll kT$) does not hold for the high frequency lines
studied in this work. For the R-J approximation to be true, we require $T\gg300$ K,
which is much higher temperature than what we expect to trace with our observations.

The background radiation $J_{\rm bg}$ 
is a composite between the cosmic microwave background radiation (CMB), as a blackbody function at 2.73 K, and the diluted infrared radiation remitted by dust. That is:

\begin{equation}
 J_{\rm bg}=J_{\nu}(2.73)+\tau_dJ_{\nu}(T_d),
\end{equation}

\noindent
where $\tau_d$ is the effective optical depth of the warm surface layer, adopted from Hollenbach \etal\ (1991), and it is defined as $\tau_d=\tau_{100\mu{\rm m}}(100\mu{\rm m}/\lambda)$.  For M17 SW we adopted an emission optical depth at 100 $\mu$m of $\tau_{100\mu {\rm m}}=0.106$ and the average dust temperature $T_d=50$ K from M92. We tried both, with and without the dust contribution to the background radiation, and we found that the contribution of the radiation by dust continuum emission is negligible at frequencies of the order of 690 GHz and 810 GHz. Nevertheless, all the following analysis include the dust contribution for completeness.

For extended (resolved) sources like the clumps in M17 SW, the radiation temperature is well estimated by the observed main beam brightness temperature $T_{mb}$. Hence, we use that quantity in the following analysis. From the LTE approximation we can assume that the excitation temperatures $T_{ex}$ of \twco and \thco $J = 6\rightarrow5$ are the same, although the terms $J_{\nu}(T_{ex})$ are not exactly the same because of the slightly ($\sim4$\%) different frequencies of the \twco and \thco lines. So, from equation (1) the ratio between \twco and \thco can be approximated as:

\begin{equation}
 \frac{T_{mb}(^{12}{\rm CO}~J=6-5)}{T_{mb}(^{13}{\rm CO}~J=6-5)}\approx\frac{1-e^{-\tau(^{12}{\rm CO}~J=6-5)}}{1-e^{-\tau(^{13}{\rm CO}~J=6-5)}},
\end{equation}

\begin{figure}[!ht]
  
  \hfill\includegraphics[width=7cm,angle=-90]{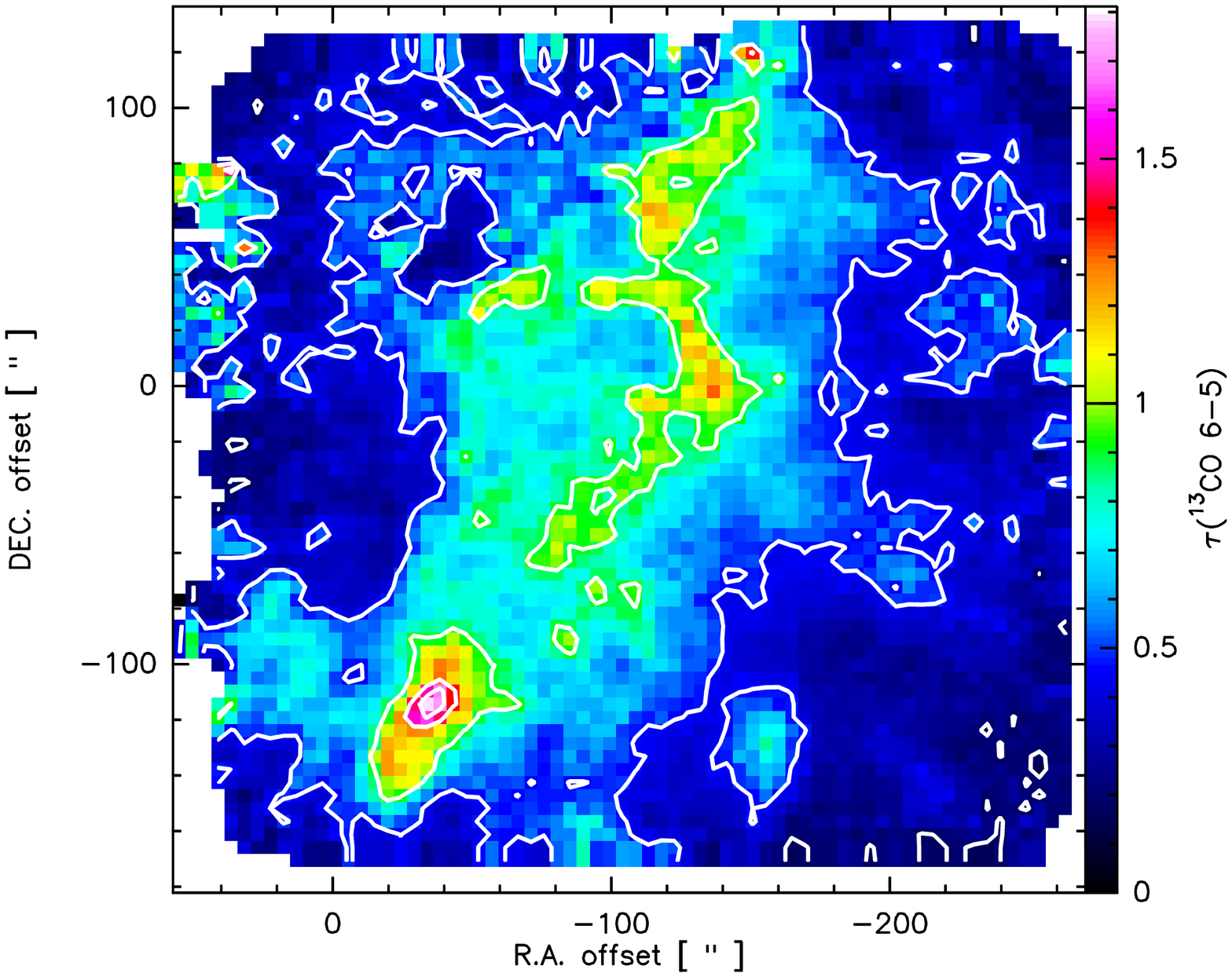}\hspace*{\fill}\\

  \hspace*{\fill}\includegraphics[width=7cm,angle=-90]{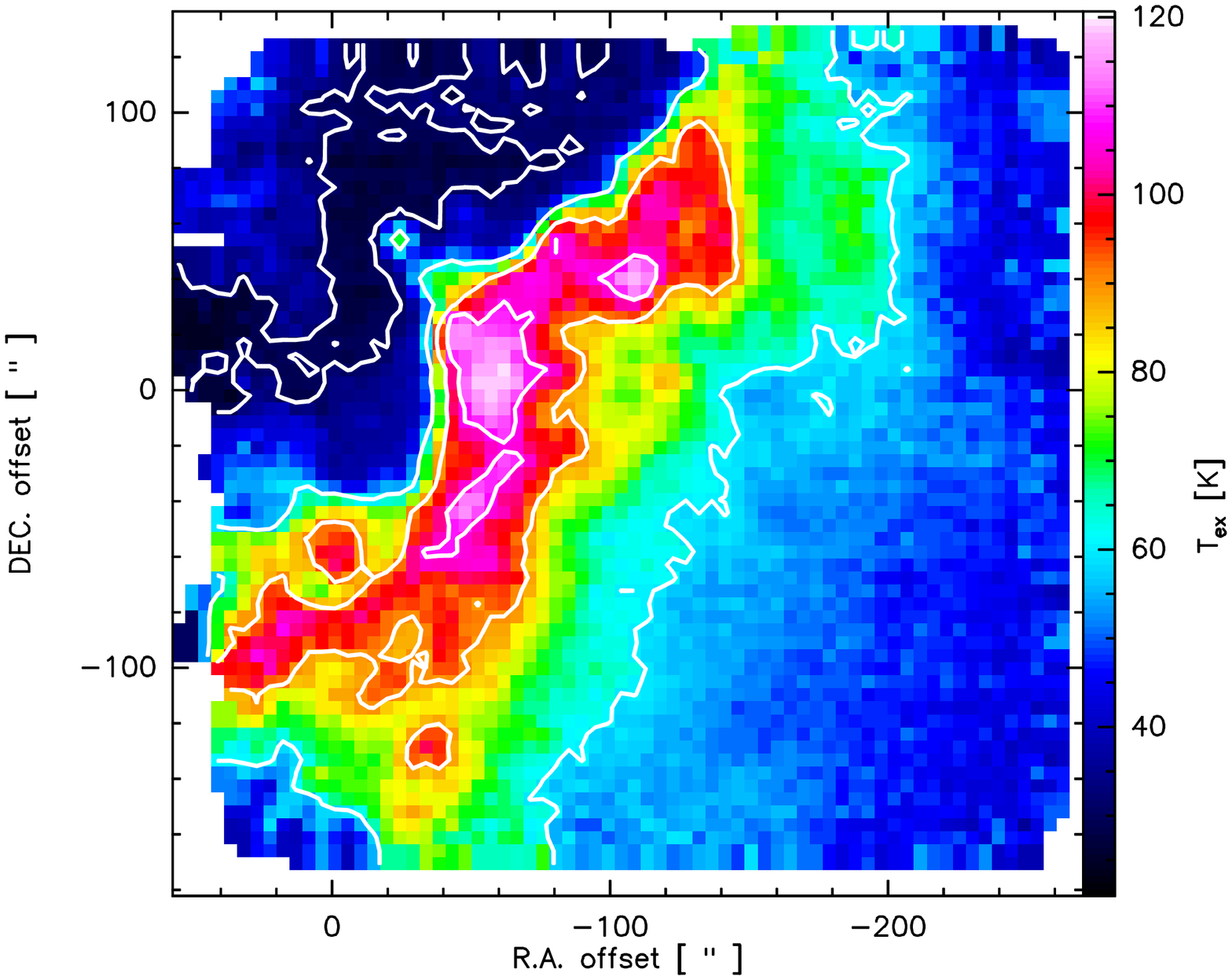}\hspace*{\fill}\\

 
  \caption{\footnotesize{LTE approximation of the optical depth (\textit{Top}) of the \thco~$J = 6\rightarrow5$ and the excitation temperature (\textit{Bottom}) of the \thco and \twco $J = 6\rightarrow5$ lines in M17 SW. Contours are the 10\%, 25\%, 50\%, 75\% and 90\% of the peak value, which is 1.9 for $\tau(\rm ^{13}CO~6-5)$ and 120 K for $T_{ex}$.}}

  \label{fig:set2-map}
\end{figure}

Following the work by Wilson \etal\ (1999), we adopted a constant [\twco/\thco] abundance 
ratio of 50 for M17 SW, which is approximately the value measured at a similar Galactic radius 
towards the W51 region (Langer \& Penzias 1990). Assuming that the optical depth is proportional
to the total column density of the molecules and, hence, to the abundance ratio between them,
we can estimate that $\tau(^{12}{\rm CO})\approx50\tau(^{13}{\rm CO})$. The \thco line is usually
optically thin, so $\tau(^{13}{\rm CO})$ could be taken out of the exponential in equation (3), and 
estimated directly. However, we do not really know if this holds for the entire M17 SW region, so
we do not apply further approximations and we solved equation (3) for $\tau(^{13}{\rm CO})$ with a 
numerical method (Newton-Raphson).

The \textit{top panel} in Figure~\ref{fig:set2-map} shows the $\tau(^{13}{\rm CO})$ map. The \thco 
line is optically thin in most of the region, with some optically thick spots (e.g., $\Delta\alpha=-30, \Delta\delta=-110$). Knowing $\tau(^{13}{\rm CO})$ we can estimate $T_{ex}$ from equation (2) using either tracer, considering that the $T_{ex}$ estimated using \twco is just $\sim0.6$\% higher than that estimated using \thco. 
The $T_{ex}$ map is shown in the \textit{bottom panel} of Figure~\ref{fig:set2-map}. This map indicates
that the warmest gas is located along the ridge of the cloud, close to the ionization front.
The temperature in this region ranges between 40 and 120 K, and the peak temperature is located 
at around ($\Delta\alpha=-60, \Delta\delta=10$). If we consider only the gas with temperatures $\ge80$ K, the warm gas would be confined to a zone of about $40''$ ($\sim0.44$ pc) next to, and along, the HII region, which agrees with previous results found by Graf \etal\ (1993).
If the gas were thermalized, then this can be the actual
map of the kinetic temperature of the gas. Otherwise, the $T_{ex}$ map can be considered as a lower 
limit of $T_K$. Since in velocity space the clumps cover the whole beam, this would imply that the \twco and \thco molecules are subthermal in the $J = 6\rightarrow5$ transition. That is, the density of the gas and the column density of \twco and \thco may be insufficient to thermalize these transitions. A more detailed analysis is presented in the next section.

\subsection{Ambient condition at selected positions (non-LTE)}

Figure~\ref{fig:set1-spec} shows the spectra of all the observed lines extracted at two different positions in the map. The \textit{top panel} shows the spectra observed at position A ($\Delta\alpha=-70'', \Delta\delta=+32''$), close to the peak emission of the \twco lines. All the lines show a double component structure with the secondary component peaking at $\sim25$ \kms. 
The \textit{middle top panel} shows the spectra at position B ($\Delta\alpha=-70'', \Delta\delta=-82''$), where the velocity integrated temperature corresponds to about 50\% of the peak emission. Here only the \twco $J = 7\rightarrow6$ line seems to have a deep at the line center. However, because the low S/N in the high frequency band, this dip may be likely due to noise.
The \textit{middle bottom panel} shows the spectra at position C ($\Delta\alpha=-60'', \Delta\delta=-30''$), which corresponds to the peak of the NE--SW strip scan reported in S88 and Graf \etal\ (1993), with beams of $40''$ and $8''$, respectively. 
The \textit{bottom panel} shows the spectra at position D ($\Delta\alpha=-100'', \Delta\delta=0''$), which is close to the continuum far-IR peak, also reported in S88.
Since we do not have dedicated observations at these positions, we extracted the spectra from the nearest pixels in our maps, convolved to the largest beam ($9.4''$) of the \thco $J = 6\rightarrow5$ line. So the spectra shown in Figure~\ref{fig:set1-spec} are the convolved spectra centered within $\pm1''$ of the indicated coordinates. This is justified because we have oversampled data.

Table~\ref{tab:gaussian-fit} shows the Gaussian fits of the spectra obtained at the four selected positions. Two Gaussian components were needed to fit the lines, except at position B, where only one component was used. The main components of the \twco lines have a line width that is about 8--9~\kms\ at position A, while the \thco has a line width of about 3~\kms\ narrower. The \ci line is the narrowest line, with a line width of $\sim4$~\kms. At position B, the \twco lines are the widest of the four lines with about 8~\kms\, and the \thco and \ci lines have about half the line width of the \twco lines. At position C and D the Gaussian parameters of the \twco $J = 7\rightarrow6$ presented uncertainties of $\sim50\%$ when let free in the fitting. However, because the line shape of the \twco $J = 7\rightarrow6$ and $J = 6\rightarrow5$ transitions are very similar, we set the line width of the $J = 7\rightarrow6$ transition to the value found for the $J = 6\rightarrow5$ line. The line width of the main components of the \twco lines at position C and D are $\sim6$\kms. That is, about 2 \kms\ narrower than the lines observed at positions A and B. This difference can be due to a higher optical depth towards the latter positions, or to the contribution of few fast-moving cloudlets (Martin, Sanders \& Hills 1984; Graf \etal\ 1993).

The \twco $\frac{7-6}{6-5}$ line ratio between the peak main beam temperatures $T_{mb}$ obtained from the Gaussian fit of the main components is $1.02\pm0.05$ at position A, $0.95\pm0.05$ at position B, $1.00\pm0.05$ at position C, and $0.99\pm0.07$ at position D.
From line ratios we can estimate the ambient conditions for these particular positions. 
We have used the non-LTE radiative transfer code RADEX\footnote{http://www.sron.rug.nl/$\sim$vdtak/radex/radex\_manual.pdf} (Van der Tak \etal~\cite{vdtak07}) for estimating the average ambient conditions (kinetic temperature, density and column density) of the molecular gas. We assumed collisional excitation by molecular hydrogen. We also assumed an homogeneous spherical symmetry in the clumps for the escape probability formalism. The collision rates between \twco and ortho- and para-H$_2$ are taken from Wernli \etal\ (2006), and can be found in the LAMDA database (Sch\"oier \etal\ 2005).
As in the LTE case, we used the cosmic microwave background radiation at 2.73 K, and we also tested the non-LTE model with and without the infrared radiation remitted by dust (eq.3) as the background source. It was also found that dust continuum emission produces a negligible effect in the non-LTE model at the frequencies of the \twco $J = 6\rightarrow5$ and $J = 7\rightarrow6$ lines. We explored molecular hydrogen densities between $10^4~\3cm$ and $10^7~\3cm$, temperatures between 5 K and 500 K, and \twco column densities between $10^{10}~\2cm$ and $10^{18}~\2cm$.

Figure~\ref{fig:ex-map-AB} shows the possible ambient conditions required to reproduce the \twco $\frac{7-6}{6-5}$ line ratios, and the peak $T_{mb}$ of the \twco~$J = 6\rightarrow5$ line observed at position A (\textit{top panel}) and B (\textit{bottom panel}). A wide range of temperatures (100 -- 450 K) and densities ($>3\times10^4~\3cm$) are possible solutions for a \twco column density per line width $N({\rm ^{12}CO})/\Delta V\sim5\times10^{16}$ \ndv.

\begin{figure}[!pt]
  
  \hfill\includegraphics[height=8cm,angle=-90]{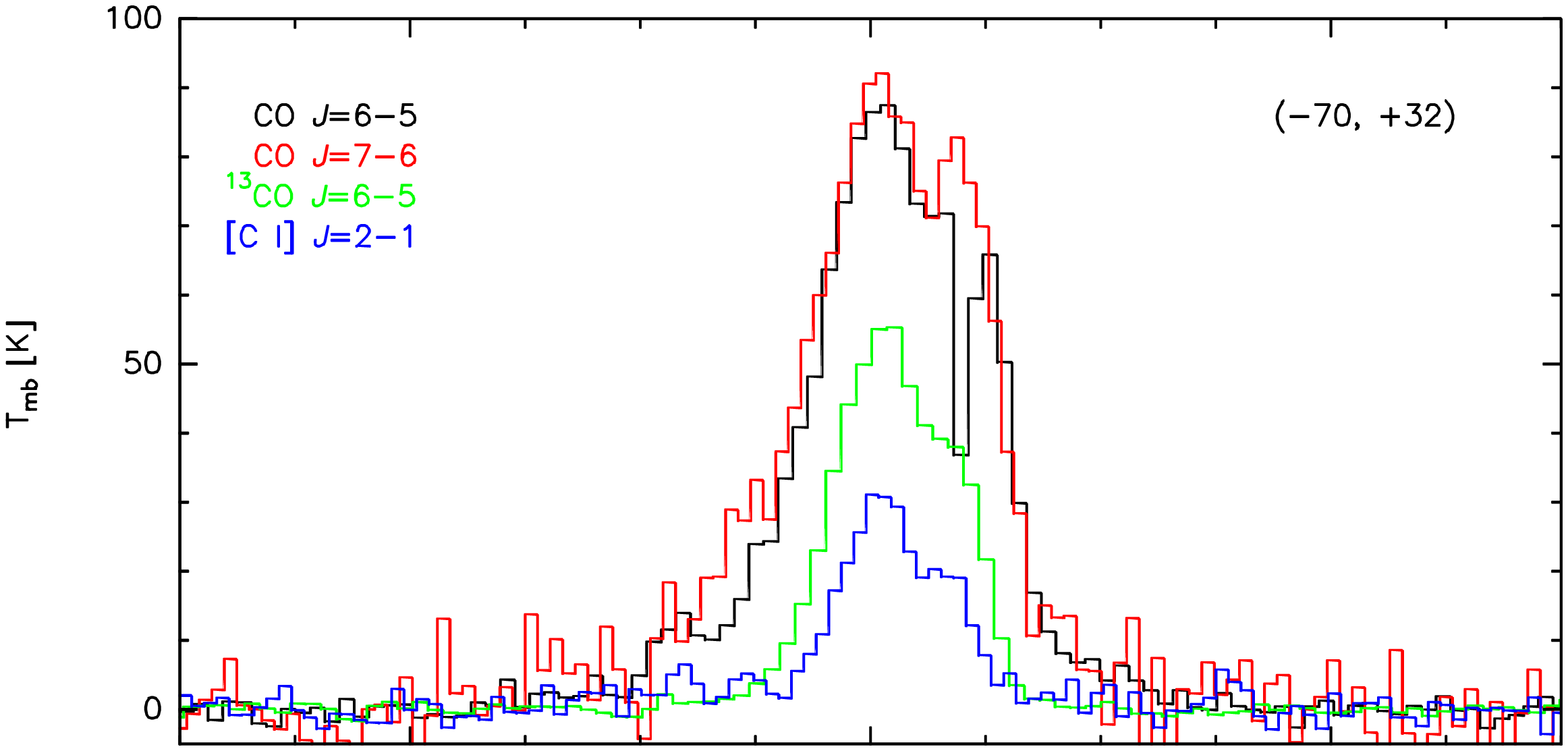}\hspace*{\fill}\\
  \vspace{-0.75cm}\\

  \hfill\includegraphics[height=8cm,angle=-90]{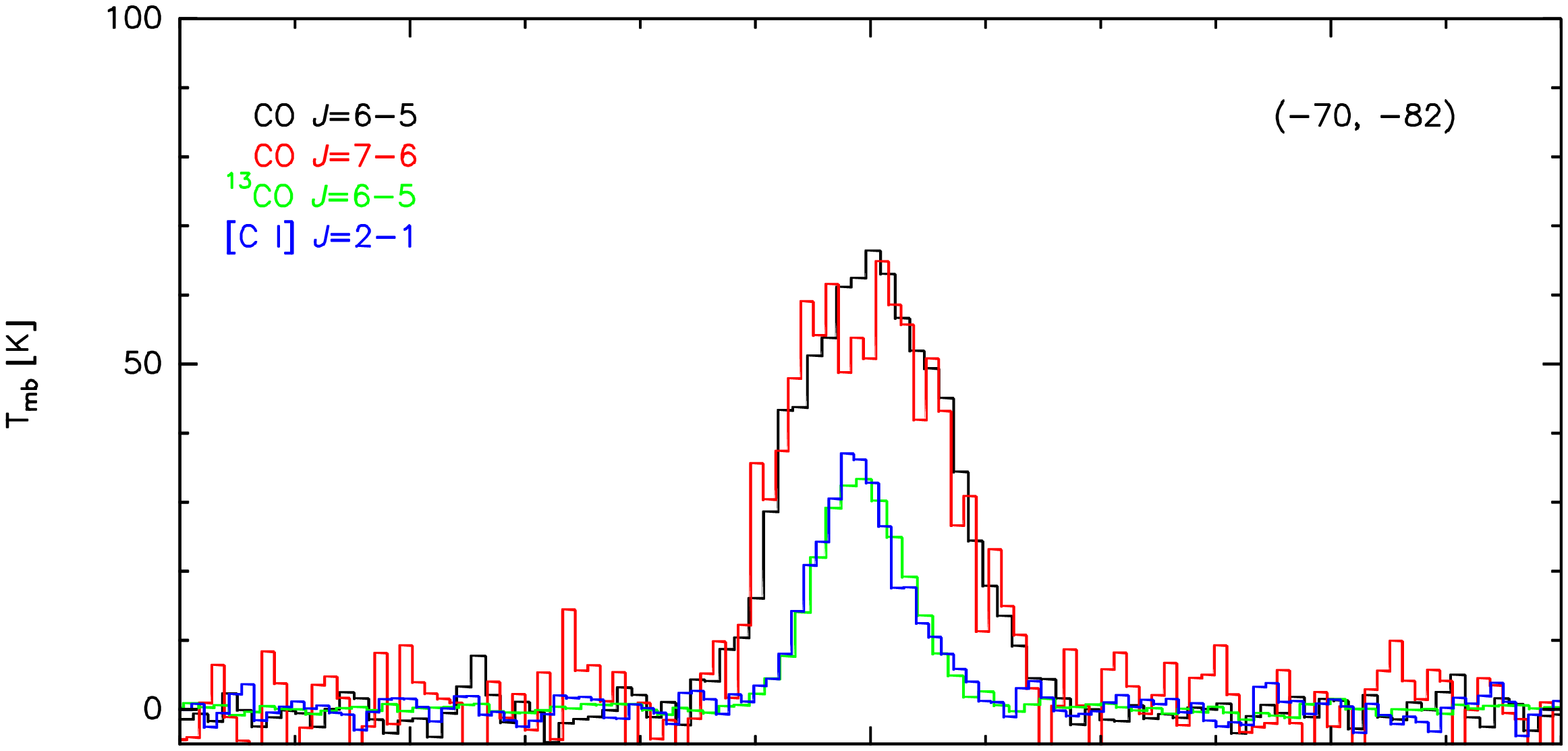}\hspace*{\fill}\\
  \vspace{-0.75cm}\\
  
  \hfill\includegraphics[height=8cm,angle=-90]{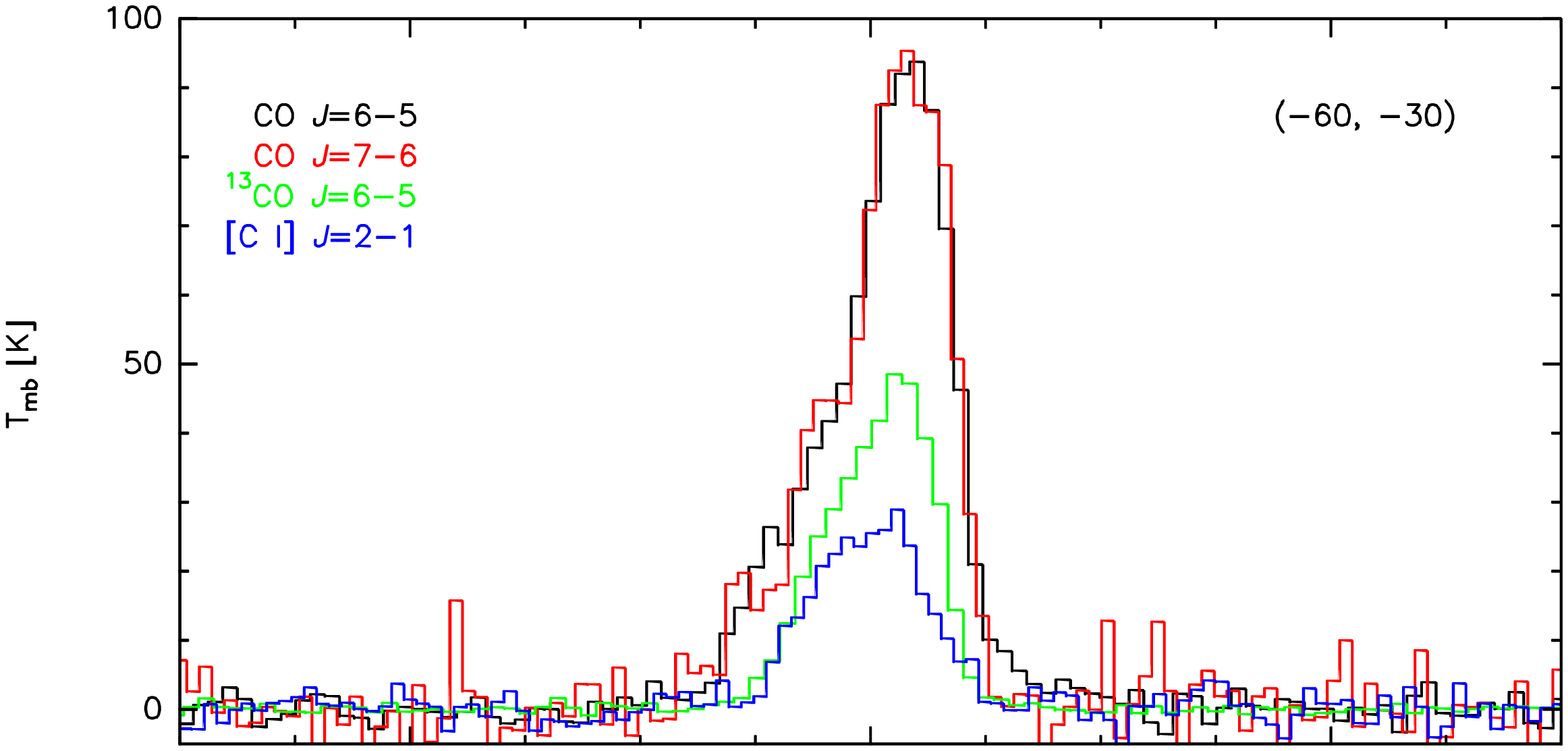}\hspace*{\fill}\\
  \vspace{-0.75cm}\\  
  
  \hfill\includegraphics[height=8cm,angle=-90]{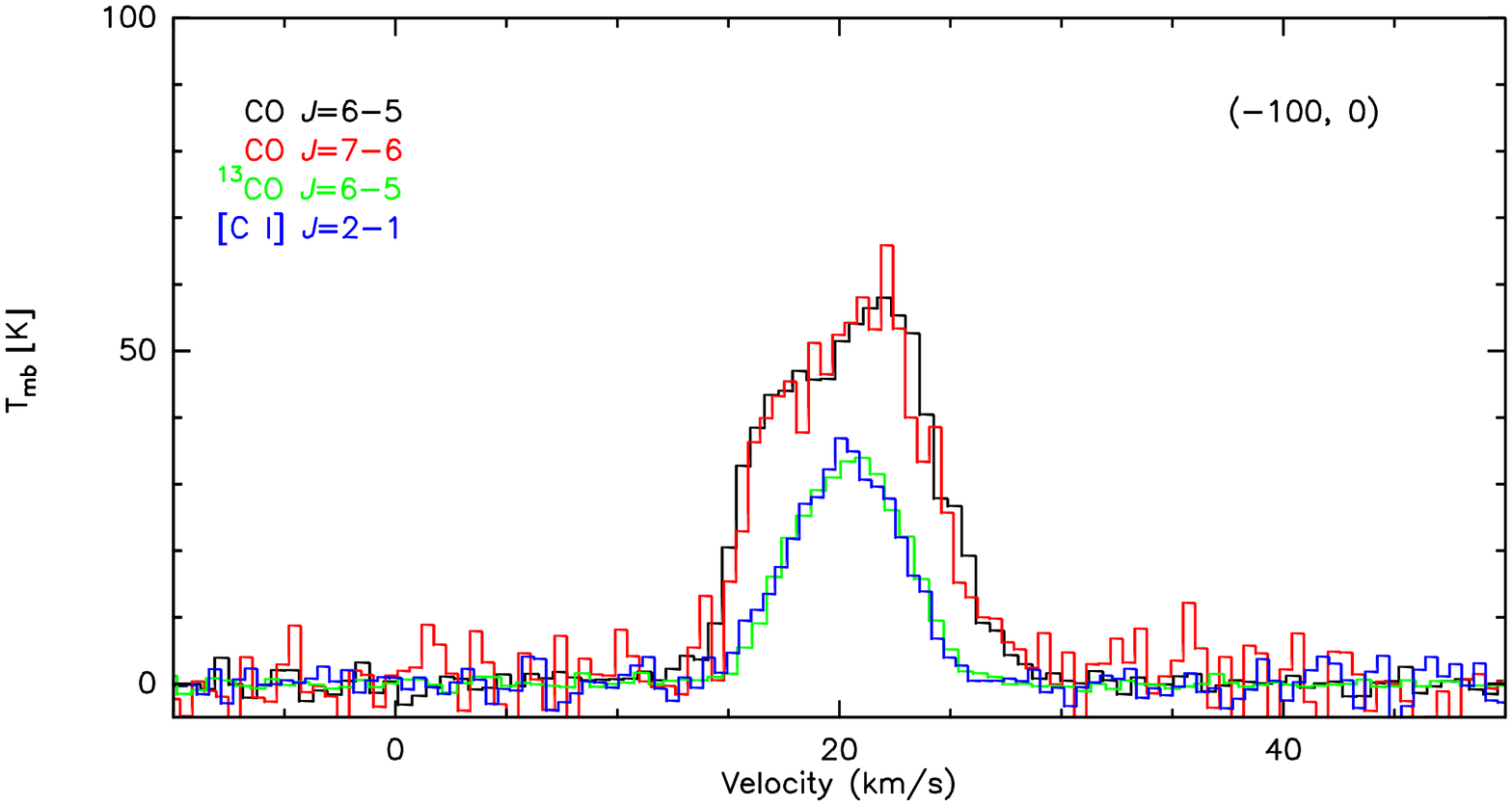}\hspace*{\fill}\\

  \caption{\footnotesize{\textit{Top} - Spectra of the four lines observed in M17 SW at position A ($\Delta\alpha=-70'', \Delta\delta=+32''$), close to the peak emission in the \twco and \thco maps. \textit{Middle top} - Spectra observed at position B ($\Delta\alpha=-70'', \Delta\delta=-82''$), where the integrated line temperatures are about 50\% of the peak emission. \textit{Middle bottom} - Spectra observed at position C ($\Delta\alpha=-60'', \Delta\delta=-30''$), the peak of the NE--SW strip scan. \textit{Bottom} - Spectra observed at position D ($\Delta\alpha=-100'', \Delta\delta=0''$), close to the continuum far-IR peak.}}

  \label{fig:set1-spec}
\end{figure}

   \begin{table}[!pt]
      \caption[]{M17 SW line parameters derived from Gaussian fits, at four selected positions.}
         \label{tab:gaussian-fit}
         \centering
         \begin{tabular}{lccccc}
            \hline\hline
	    \noalign{\smallskip}
            Molecule--$J$ & $I_{mb}$ & $V$    & $T_{mb}$ & $\Delta V$ \\
                          & [K \kms] & [\kms] &     [K]    &   [\kms] \\
            \noalign{\smallskip}
            \hline
            \noalign{\smallskip}
            \multicolumn{5}{c} {Position A $(-70'', +32'')$} \\
            \noalign{\smallskip}
            \hline
            \noalign{\smallskip}

            \multirow{2}{*}{\twco~$J$=6--5} & 723.1$\pm$7.5 & 20.4$\pm$0.04 & 82.9$\pm$1.5 & 8.2$\pm$0.11\\
		                            & ~44.6$\pm$3.3 & 25.1$\pm$0.03 & 31.5$\pm$2.9 & 1.3$\pm$0.07\\

	    \noalign{}\\
            \multirow{2}{*}{\twco~$J$=7--6} & 825.4$\pm$24.1 & 20.3$\pm$0.13 & 84.6$\pm$3.6 & 9.2$\pm$0.28\\
		                            & ~56.9$\pm$12.8 & 24.2$\pm$0.15 & 26.5$\pm$7.3 & 2.0$\pm$0.32\\
            
	    \noalign{}\\
            \multirow{2}{*}{\thco~$J$=6--5} & 301.2$\pm$4.2 & 20.2$\pm$0.04 & 54.7$\pm$1.1 & 5.2$\pm$0.08\\
		                            & ~51.3$\pm$3.6 & 23.9$\pm$0.04 & 21.2$\pm$1.7 & 2.3$\pm$0.09\\

	    \noalign{}\\
            \multirow{2}{*}{{\rm [C I]}~$J$=2--1} & 138.5$\pm$11.5 & 20.2$\pm$0.16 & 30.0$\pm$3.8 & 4.3$\pm$0.42\\
		                                  & ~29.9$\pm$10.1 & 23.6$\pm$0.18 & 12.7$\pm$5.2 & 2.2$\pm$0.51\\

            \noalign{\smallskip}
            \hline
	    
            \noalign{\smallskip}
            \multicolumn{5}{c} {Position B $(-70'', -82'')$} \\
            \noalign{\smallskip}
            \hline
            \noalign{\smallskip}

		\twco~$J$=6--5 & 537.4$\pm$6.9 & 19.7$\pm$0.05 & 65.7$\pm$1.3 & 7.7$\pm$0.11\\
            
	    \noalign{}\\
		\twco~$J$=7--6 & 528.5$\pm$16.1 & 19.5$\pm$0.12 & 62.2$\pm$2.8 & 7.9$\pm$0.26\\
            
	    \noalign{}\\
		\thco~$J$=6--5 & 166.7$\pm$1.6 & 19.3$\pm$0.02 & 33.5$\pm$0.5 & 4.7$\pm$0.05\\

	    \noalign{}\\
		{\rm [C I]}~$J$=2--1 & 169.7$\pm$4.2 & 19.2$\pm$0.06 & 34.4$\pm$1.4 & 4.6$\pm$0.14\\
			
            \noalign{\smallskip}
            \hline
	    
            \noalign{\smallskip}
            \multicolumn{5}{c} {Position C $(-60'', -30'')$} \\
            \noalign{\smallskip}
            \hline
            \noalign{\smallskip}
	    
            \multirow{2}{*}{\twco~$J$=6--5} & 220.4$\pm$3.1 & 17.9$\pm$0.04 & 33.7$\pm$0.6 & 6.1$\pm$0.06\\
		                            & 365.8$\pm$4.9 & 21.9$\pm$0.03 & 86.2$\pm$1.6 & 3.9$\pm$0.05\\

	    \noalign{}\\
            \multirow{2}{*}{\twco~$J$=7--6} & 220.2$\pm$16.8 & 17.9$\pm$0.31 & 33.6$\pm$2.6 & 6.1$^{\mathrm{a}}$\\
		                            & 367.1$\pm$16.4 & 21.9$\pm$0.08 & 86.5$\pm$4.0 & 3.9$^{\mathrm{a}}$\\
            
	    \noalign{}\\
            \multirow{2}{*}{\thco~$J$=6--5} & 145.7$\pm$11.5 & 18.9$\pm$0.17 & 29.8$\pm$2.7 & 4.6$\pm$0.21\\
		                            & 120.3$\pm$11.3 & 21.8$\pm$0.05 & 36.9$\pm$3.6 & 3.1$\pm$0.09\\

	    \noalign{}\\
            \multirow{2}{*}{{\rm [C I]}~$J$=2--1} & ~41.6$\pm$27.6 & 17.5$\pm$0.55 & 12.1$\pm$8.6 & 3.2$\pm$0.77\\
		                                  & 124.5$\pm$28.6 & 20.7$\pm$0.48 & 24.9$\pm$7.3 & 4.5$\pm$0.72\\
		
            \noalign{\smallskip}
            \hline	    	    
	    
            \noalign{\smallskip}
            \multicolumn{5}{c} {Position D $(-100'', 0'')$} \\
            \noalign{\smallskip}
            \hline
            \noalign{\smallskip}
	    
            \multirow{2}{*}{\twco~$J$=6--5} & 136.1$\pm$10.9 & 16.8$\pm$0.09 & 34.6$\pm$3.2 & 3.7$\pm$0.17\\
		                            & 374.4$\pm$11.6 & 21.8$\pm$0.09 & 57.9$\pm$2.5 & 6.1$\pm$0.18\\

	    \noalign{}\\
            \multirow{2}{*}{\twco~$J$=7--6} & 114.1$\pm$11.5 & 16.9$\pm$0.19 & 29.1$\pm$3.2 & 3.7$^{\mathrm{a}}$\\
		                            & 369.7$\pm$18.9 & 21.6$\pm$0.15 & 57.2$\pm$3.4 & 6.1$^{\mathrm{a}}$\\
            
	    \noalign{}\\
            \multirow{2}{*}{\thco~$J$=6--5} & ~82.7$\pm$4.2 & 18.4$\pm$0.09 & 20.0$\pm$1.3 & 3.9$\pm$0.15\\
		                            & 131.3$\pm$4.1 & 21.6$\pm$0.05 & 28.7$\pm$0.9 & 4.2$\pm$0.04\\

	    \noalign{}\\
            \multirow{2}{*}{{\rm [C I]}~$J$=2--1} & ~95.9$\pm$5.2 & 18.7$\pm$0.17 & 17.9$\pm$1.5 & 5.0$\pm$0.30\\
		                                  & 120.9$\pm$1.5 & 21.4$\pm$0.06 & 24.5$\pm$1.3 & 4.6$\pm$0.24\\
		
            \noalign{\smallskip}
            \hline	    		    

         \end{tabular}
\begin{list}{}{}
\item[$^{\mathrm{a}}$] The uncertainty of this parameter was larger than 50\%, when let free in the Gaussian fitting. So we set its value to the one found for the corresponding Gaussian component of the \twco $J$=6--5 line.
\end{list}     
   \end{table}

\begin{figure}[!ht]
  \hfill\includegraphics[width=8cm]{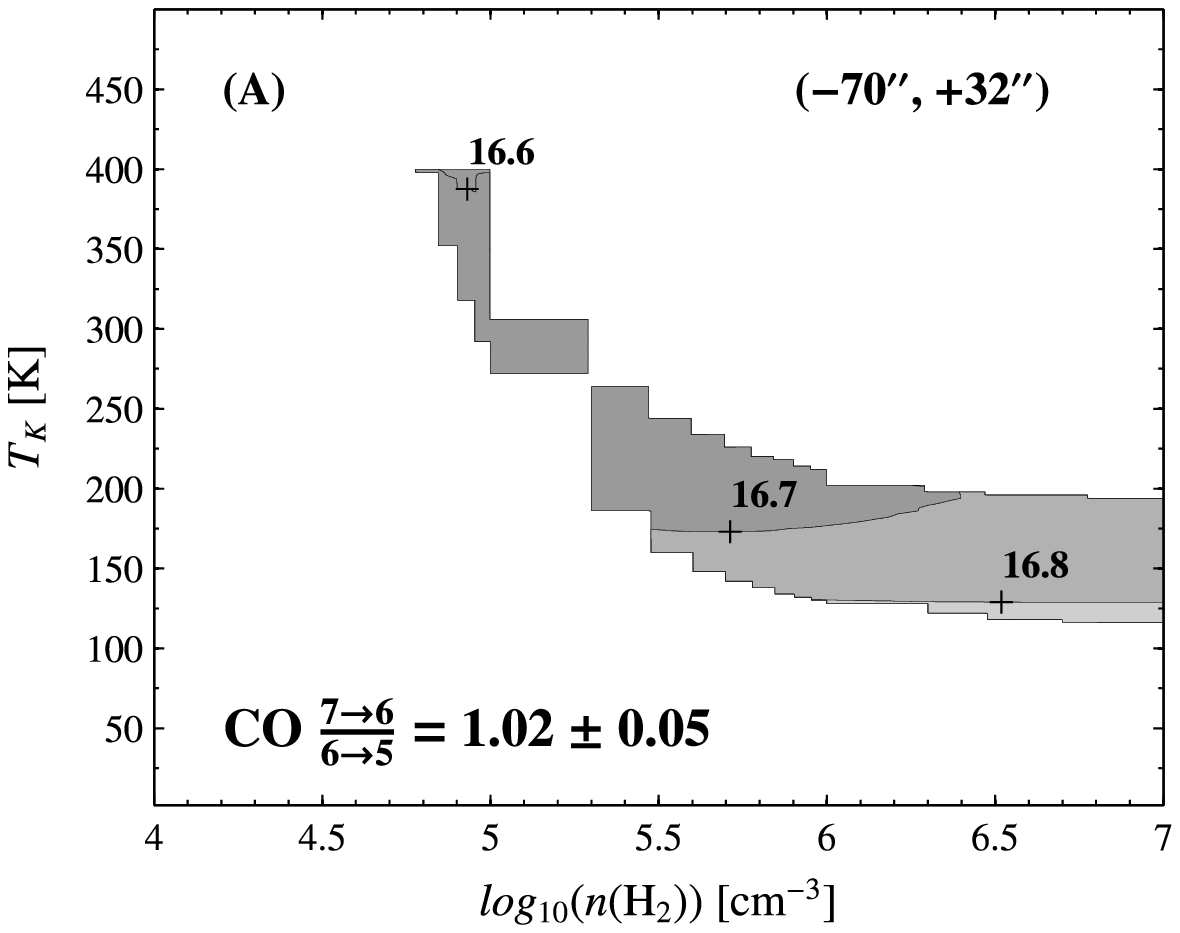}\hspace*{\fill}\\

  \hfill\includegraphics[width=8cm]{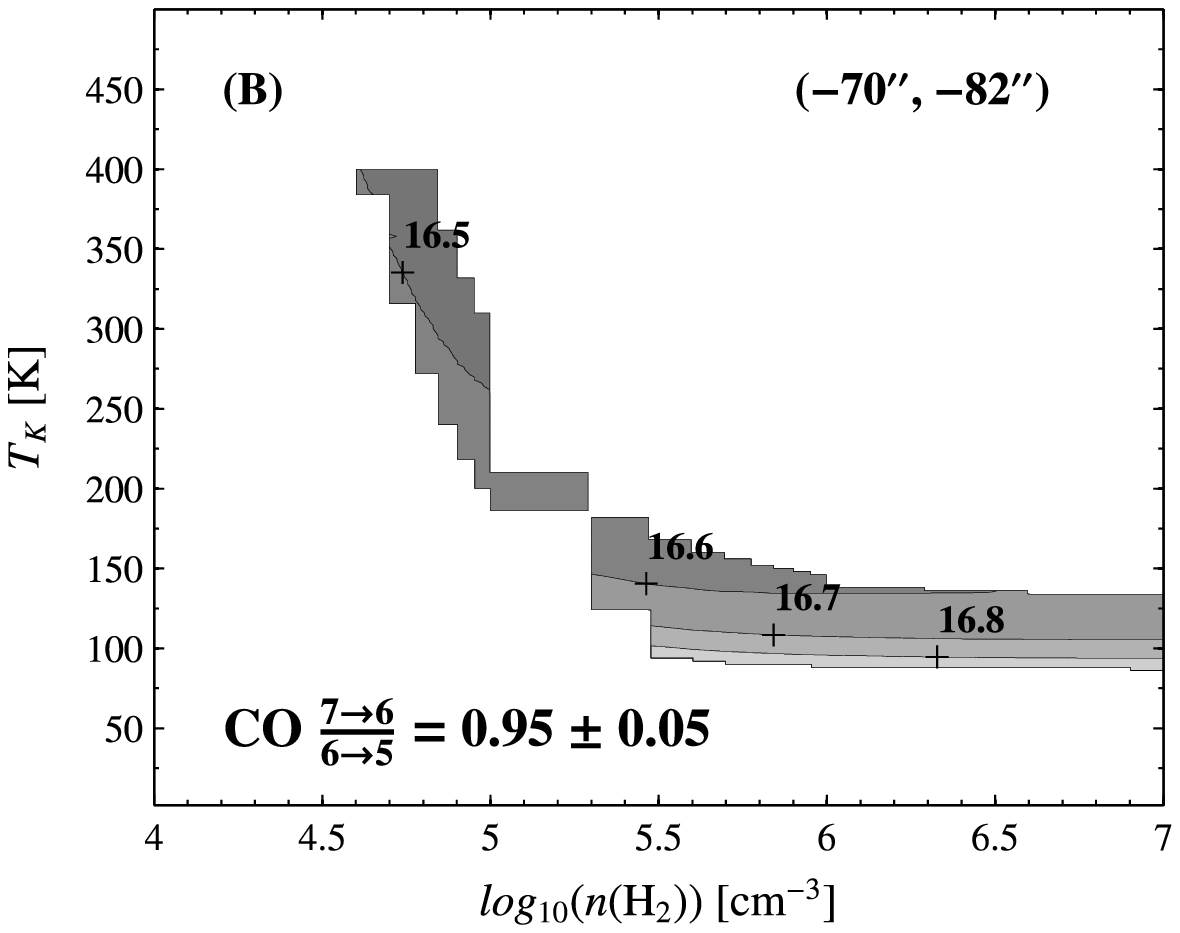}\hspace*{\fill}\\

  \caption{\footnotesize{\textit{Top} - The gray scale and contours represent the average ($log_{10}$ scale) column density per line width (\ndv) 
required to reproduce the observed \twco~$\frac{7-6}{6-5}$ line ratio between the peak main beam temperatures $T_{mb}$ and the the peak $T_{mb}$ of the \twco~$J=6\rightarrow5$ line observed at position A ($\Delta\alpha=-70'', \Delta\delta=+32''$), for different kinetic temperatures $T_K$ and densities $n(\rm H_2)$.
\textit{Bottom} - Same as top, but at position B ($\Delta\alpha=-70'', \Delta\delta=-82''$).}}

  \label{fig:ex-map-AB}
\end{figure}

\begin{figure}[!ht]
  \hfill\includegraphics[width=8cm]{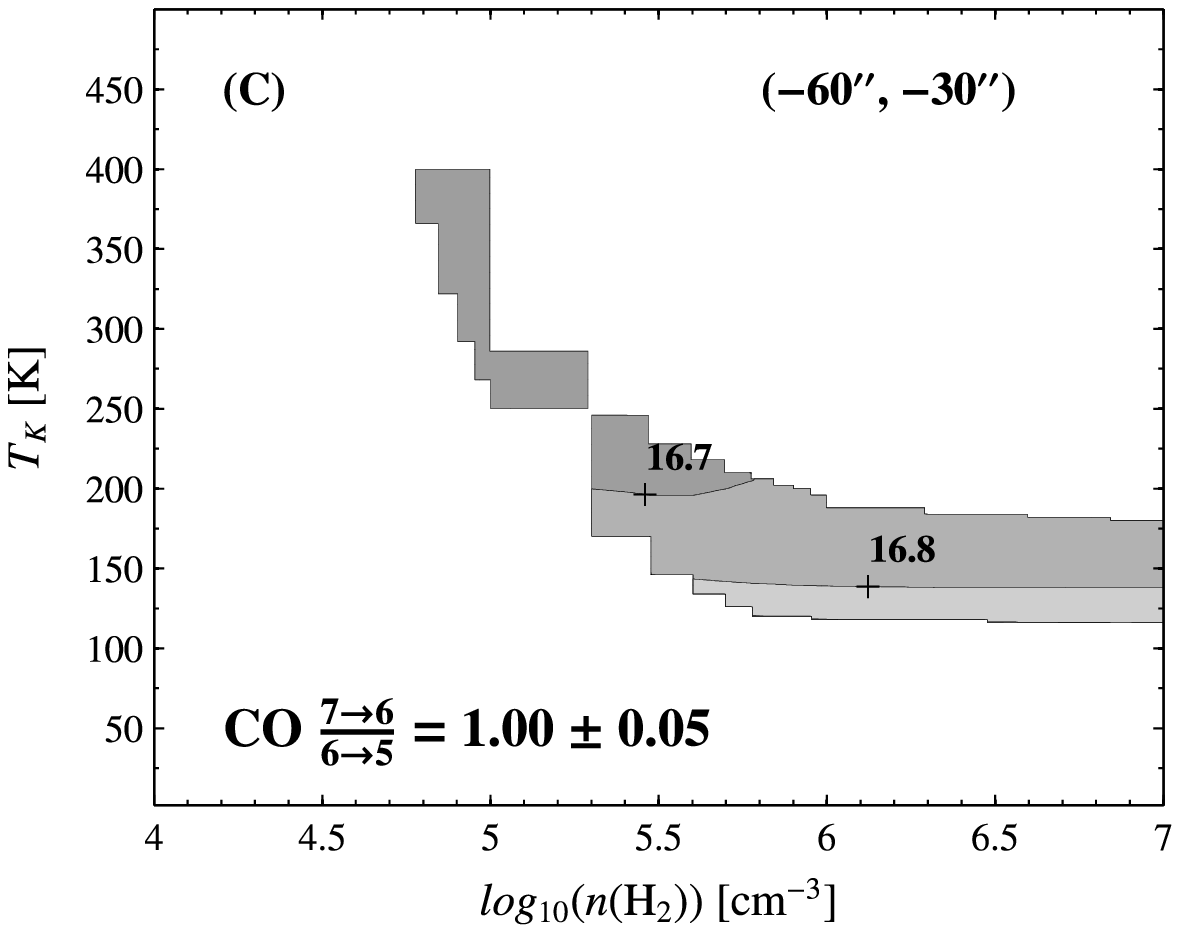}\hspace*{\fill}\\

  \hfill\includegraphics[width=8cm]{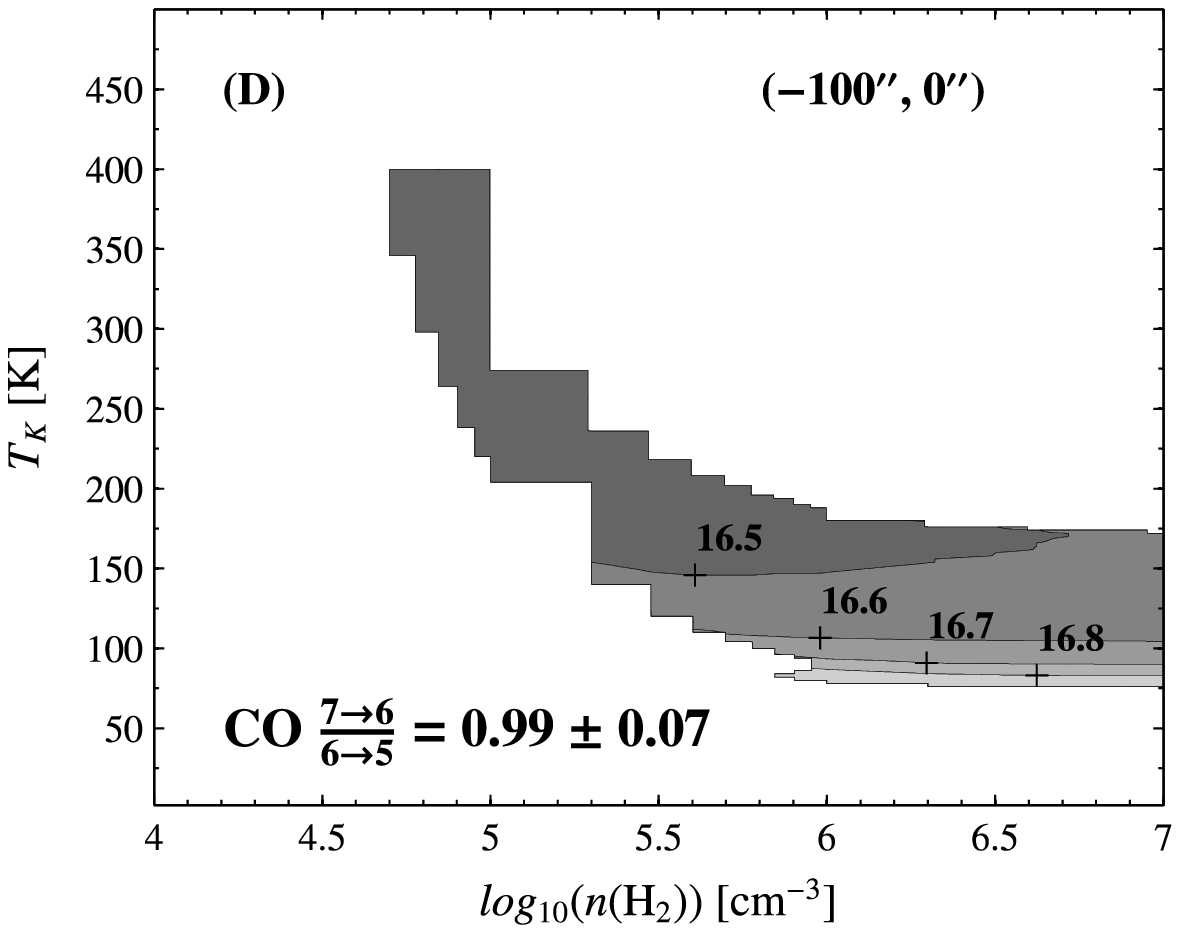}\hspace*{\fill}\\

  \caption{\footnotesize{\textit{Top} - The gray scale and contours represent the average ($log_{10}$ scale) column density per line width (\ndv) 
required to reproduce the observed \twco~$\frac{7-6}{6-5}$ line ratio between the peak main beam temperatures $T_{mb}$ and the the peak $T_{mb}$ of the \twco~$J=6\rightarrow5$ line observed at position C ($\Delta\alpha=-60'', \Delta\delta=-30''$), for different kinetic temperatures $T_K$ and densities $n(\rm H_2)$.
\textit{Bottom} - Same as top, but at position D ($\Delta\alpha=-100'', \Delta\delta=0''$).}}

  \label{fig:ex-map-CD}
\end{figure}

Figure~\ref{fig:ex-map-CD} shows the possible ambient conditions estimated for position C (\textit{top panel}) and D (\textit{bottom panel}). The combinations of temperatures and densities required to reproduce the line ratios and peak temperatures are similar to those found for position A and B. Although, the range of possible temperatures (for a given density) at position D is larger than at the other positions. The column densities differ due to the different line strengths observed at the four positions (Table~\ref{tab:gaussian-fit}).

In order to constrain the range of solutions we can adopt the average $5\times10^5~\3cm$ density estimated by M92, which is also similar to the mean density of the clumps estimated by SG90. This is a sensitive assumption for a collision dominated scenario since this density is larger than the critical density of both \twco lines for $T_K\ge20$ K.
However, at this density ($5\times10^5~\3cm$) the temperature cannot be higher than 230 K in order to reproduce the line ratio and the peak $T_{mb}$ of the \twco $J = 6\rightarrow5$ line observed at position A. And it cannot be higher than 150 K at position B. At position C the limit is about 220 K, and at position D it is about 200 K.
These are lower kinetic temperatures than the 1000 K estimated for the dense clumps in the three component model proposed by M92. Our upper limits for the kinetic temperature agree with the results reported in previous work (e.g Harris \etal\ 1987; S88; SG90).
From the map of the excitation temperature $T_{ex}$ estimated from the LTE model (Figure~\ref{fig:set2-map}) the lower limits for $T_K$ would be $\sim110$ K and $\sim80$ K at position A \& C and B \& D, respectively. These are similar (within 30\%) to the lowest temperatures obtained with the non-LTE models (Figures~\ref{fig:ex-map-AB} \& \ref{fig:ex-map-CD}).

According to the radiative transfer model temperatures up to 400 K, and higher, are also possible, but they require densities $<10^5~\3cm$ in order to reproduce the observed line ratios and peak temperatures. These densities and temperatures are consistent with the estimates made based on previous observations of the \twco $J = 7\rightarrow6$ and $J = 14\rightarrow13$ lines (Harris \etal\ 1987; SG90). On the other hand, clumps with densities $>10^6~\3cm$ could also reproduce the observed ratios and peak $T_{mb}$ in all the positions, at temperatures $\le200$ K. However, these would be at the lower limit of the temperature range estimated in Harris \etal\ (1987) and SG90. The densities and temperatures found for M17 SW are similar to those found in W3 Main (Kramer \etal\ 2004), but higher (although compatible) than the kinetic temperatures found in Carina, where lower limits between 30 K and 50 K were estimated (Kramer \etal\ 2008).

\subsubsection{Column densities at selected positions}

The column density per line width $N({\rm ^{12}CO})/\Delta V$ at position A varies over a small range of $4-6\times10^{16}~\ndv$. If we consider an average line width of 8.7 \kms\ estimated for the \twco $J = 6\rightarrow5$ and $J = 7\rightarrow6$ lines (Table~\ref{tab:gaussian-fit}) and the average $N({\rm CO})/\Delta V=5\times10^{16}~\ndv$, then we have a total column density of $N({\rm ^{12}CO})\approx4\times10^{17}~\2cm$. 


The model indicates that at position B the \twco column density per line width would be $\sim4\times10^{16}~\ndv$. The average line width of the lines at position B is $\sim7.7$ \kms\ (Table~\ref{tab:gaussian-fit}), which gives a total column density $N({\rm ^{12}CO})\approx3\times10^{17}~\2cm$, similar to the column found at position A. 
At positions C and D the average column densities per line width are $\sim8\times10^{16}~\ndv$ and $\sim5\times10^{16}~\ndv$, respectively. Considering a line width of $\sim6$~\kms\ we obtain similar column densities as in the previous two positions. That is $N({\rm ^{12}CO})\approx5\times10^{17}~\2cm$ and $N({\rm ^{12}CO})\approx3\times10^{17}~\2cm$, for positions C and D, respectively. At positions A and C the lines are optically thin, with $\tau$ ranging from about 0.6 to 1 for $T_K\ge150$ K. At temperatures $\le150$ K, the lines become optically thick ($1\le\tau\le3$). The optically thin limit at positions B and D is reached at $T_K\sim120$ K, with about the same ranges of optical depths as before, for temperatures higher or lower than 120 K.

Assuming a density of $5\times10^5~\3cm$ and average temperatures of 200 K at position A and 150 K at position B, the non-LTE model indicates that \thco column densities of $\sim1.5\times10^{17}~\2cm$ and $\sim7.4\times10^{16}~\2cm$ would be required to reproduce the observed strength (Table~\ref{tab:gaussian-fit}) of the \thco $J=6\rightarrow5$ line at these positions, respectively. For a temperature of 200 K, the \thco column density at position C and D would be $\sim5.6\times10^{16}~\2cm$. And the excitation temperatures would be $\sim180$ K at position A, C and D, and $\sim140$ K at position B, which are higher temperatures than estimated with the LTE approximation. However, these excitation temperatures are just between 10 and 20 K lower than the assumed kinetic temperatures, which indicates that this lines are close (within 10\%) to the thermal equilibrium.

From their C$^{18}$O $J=2\rightarrow1$ observations, S88 estimated a \twco beam-averaged column density of $\sim2\times10^{19}~\2cm$, considering a [\twco]/[C$^{18}$O] abundance ratio of 500. Hence, the column densities found for the four selected positions in M17 SW suggest that the warm ($T_k>100$ K) and dense ($n(\rm H_2)\ge10^4~\3cm$) gas traced by the mid-$J$ \twco lines represent $\lesssim2$\% of the bulk of the cold ($T_k<50$ K) and less dense ($n(\rm H_2)\sim10^3~\3cm$) gas traced by the low-$J$ \twco lines.

%

\subsubsection{Volume filling factors}

The clump volume filling factor $\Phi_V$ can be estimated from the ratio between the average volume density $n_{av}$ per beam, and the average clump density $n_{clump}\sim5\times10^5~\3cm$ derived from the non-LTE model (e.g. Kramer et al. 2004).
The average volume density per beam can be estimated from the total column density of the gas, and the line of sight extent of the cloud ($D_{cloud}$). That is $n_{av}\sim N(\rm H_2)/D_{cloud}$. Following the work by Howe \etal\ (2000) we can assume a \thco abundance ratio of $1.5\times10^{-6}$ relative to H$_2$, and estimate the hydrogen column densities of $N_{\rm A}(\rm H_2)\sim2.3\times10^{23}~\2cm$, $N_{\rm B}(\rm H_2)\sim1.1\times10^{23}~\2cm$, and $N_{\rm C,D}(\rm H_2)\sim8.4\times10^{22}~\2cm$, for the four selected positions.

The line of sight extent of the cloud is a difficult parameter to estimate. From a $13''$ ($\sim0.14$ pc) beam-averaged column density of $N(\rm H_2)\sim8\times10^{23}~\2cm$ a volume filling factor of 0.13 was estimated by SG90. While Howe \etal\ (2000) reported a $\Phi_V\sim0.002$ from the total column density of $N(\rm H_2)\sim4\times10^{22}~\2cm$ estimated at the peak column density of their \thco $J = 1\rightarrow0$ map, and assuming a cloud extent of 3 pc, which was deconvolved from the $4'$ beam of the SWAS space telescope. The line of sight extent should be larger than the smallest possible clump size ($\sim0.1$ pc) that we can deconvolve from our $9.4''$ beam. But we do not think it can be as large as 3 pc, which is about the size of the maps we present here. This holds at least for the region of bright \twco and \thco emission close to the ionization front, where our four selected positions are taken from. If we take the average between the upper (3 pc) and lower (0.1 pc) limits of the cloud extent, we would obtain a cloud size of $\sim1.6$ pc. This line of size extent of the cloud is uncertain, but perhaps more realistic given the geometry of M17 SW and the high resolution of our maps. Besides, it is similar to the diameter of the \ci emitting region ($\approx1$ pc) estimated by Genzel \etal\ (1988), and the narrow spatial extension ($\sim1.3$ pc) of the \thco\ $J=6\rightarrow5$ and \ci\ 370 $\mu$m lines along the strip line at P.A=$63^{\circ}$ (Figures~\ref{fig:strip-lines} \& \ref{fig:strip-lines2}).

Using the total column densities estimated for the four selected positions, and $D_{cloud}=1.6$ pc, the average volume densities at position A and B would be $\sim5.3\times10^4~\3cm$ and $\sim2.5\times10^4~\3cm$, respectively, and $\sim1.9\times10^4~\3cm$ at position C, and D. This in turn gives volume filling factors $\Phi_V=n_{av}/n_{clump}$ of $\sim0.106$, $\sim0.050$ and $0.038$ at positions A, B and C/D, respectively. These volume filling factors, as well as the total hydrogen densities estimated here, are larger than those estimated by Howe \etal\ (2000), but smaller than the ones reported in SG90. This is an expected and reasonable result since the \thco $J = 6\rightarrow5$ line traces only the warm and dense clumps and not the interclump medium. Besides, the volume filling factors estimated at the four selected positions are in close agreement to those estimated in other star forming regions using clumpy PDR models (e.g. S140, W3 Main; Spaans \& van Dishoeck 1997; Kramer \etal\ 2004).

\subsubsection{Jeans stability of the clumps}

With an average density of $5\times10^5~\3cm$, and an average clump size of 0.2 pc in diameter, which gives a typical total clump mass of $\sim100$ $M_{\sun}$ in molecular hydrogen, M92 estimated that these clumps are not in pressure equilibrium with the interclump gas (with average density $3\times10^3~\3cm$ and temperature of 200 K), but rather that they are self-gravitating. With these parameters, and a temperature of the order of 1000 K, the Jeans mass and radius should be about $1500~M_{\sun}$ and 0.3 pc, respectively. Hence these clumps are not near the collapsing regime. Even with our upper limits for the temperatures of the clumps of 230 K and 150 K at position A and B, and 220 K and 200 K at position C and D, respectively, the Jeans mass and radius of these clumps would still be larger than those estimated with the average density of $5\times10^5~\3cm$. Temperatures $<170$ K would be required to break the Jeans stability at that density. This means that the clumps at position B should have a slightly lower density of $\sim3\times10^5~\3cm$ (or lower) to be Jeans stable at a temperature of about 150 K (or higher).

\subsection{Follow-up work}

A higher resolution map of the 609 $\mu$m (492 GHz) $^3P_1\rightarrow{^3P_0}$ fine-structure transition of \ci will be obtained with FLASH on APEX, in order to constrain the ambient conditions of the interclump medium and the halo in M17 SW.
More complex radiative transfer codes like RATRAN (Hogerheijde \& van der Tak 2000) and $\beta 3D$ (Poelman \& Spaans 2005), will be used to model the internal dynamics, temperature and density structure of individual clouds. The models will also allow us to explore in detail the effect of absorbing foreground clouds, or multiple cloud components, in the line profiles. Our PDR code (Meijerink \& Spaans 2005) will provide the abundances of the molecular and atomic species, according to the UV flux estimated from historical data and our mid-$J$ lines data. All together, these models will aid to test and constrain the heating and cooling of the irradiated gas.

\section{Conclusions}

We have used the dual color heterodyne receiver array of 7 pixels CHAMP$^+$ 
on the APEX telescope, to map a region of about 2.6 pc $\times$ 2.9 pc 
in the $J = 6\rightarrow5$ and $J = 7\rightarrow6$ lines of \twco, the 
\thco $J = 6\rightarrow5$ and the $^3P_2\rightarrow{^3P_1}$ 370 $\mu$m ($J=2\rightarrow1$)
fine-structure transition of \ci in M17 SW nebula.

The completely different structure and distribution of the $^3P_2\rightarrow{^3P_1}$ 370 $\mu$m emission, and its critical density, indicate that this emission arises from the interclump medium ($\sim3\times10^3~\3cm$). On the other hand, the mid-$J$ lines of \twco and the isotope emissions, arise from the high density ($\sim5\times10^5~\3cm$) and clumpy region.

The spatial extent of the warm gas (40-230 K) traced by the \twco $J=7\rightarrow6$ line is about 2.2 pc from the ridge of the M17 SW complex, which is smaller than the extent observed in the low-$J$ \twco and C$^{18}$O lines reported in previous work. The \thco\ $J=6\rightarrow5$ and \ci\ 370 $\mu$m lines, have a narrower spatial extent of about 1.3 pc along a strip line at P.A=$63^{\circ}$.

An LTE approximation of the excitation temperature provides lower limits for the kinetic temperature. The warmest gas is located along the ridge of the cloud, close to the ionization front. In this region the excitation temperature range between 40 and 120 K.
A non-LTE estimate of the ambient conditions at four selected positions of M17 SW indicates that the high density clumps ($\sim5\times10^5~\3cm$) cannot have temperatures higher than 230 K. The warm ($T_k>100$ K) and dense ($n(\rm H_2)\ge10^4~\3cm$) gas traced at the four selected positions by the mid-$J$ \twco lines represent $\sim2$\% of the bulk of the molecular gas traced by the low-$J$ \twco lines. Volume filling factors of the warm gas ranging from 0.04 to 0.11 were found at these positions.

\acknowledgements{We are grateful with the MPfIR team and the
APEX staff for their help and support during and after the observations.
We are grateful to J. Stutzki for providing low-$J$ \twco data and to C. Brogan for providing the 21 cm map.
We thank the referee for the careful reading of the manuscript and constructive comments.
Molecular Databases that have been helpful include the NASA/JPL, LAMDA and NIST.
Construction of CHAMP$^+$ is a collaboration between the Max-Planck-
Institut f\"ur Radioastronomie Bonn, SRON Groningen, the Netherlands
Research School for Astronomy (NOVA), and the Kavli Institute of Nanoscience
at Delft University of Technology, with support from the Netherlands
Organization for Scientific Research (NWO) grant 600.063.310.10.
}


\end{document}